\documentclass[krantz1,ChapterTOCs]{krantz} 
\usepackage{fixltx2e,fix-cm}
\usepackage{amssymb}
\usepackage{amsmath}
\usepackage{graphicx}
\usepackage{subfigure}
\usepackage{makeidx}
\usepackage{multicol}
\usepackage{ulem}
\RequirePackage{tikz}
\usetikzlibrary{positioning}
\newcommand{\AR}{\mbox{\tiny{AR}}}
\newcommand{\NE}{\mbox{\tiny{NE}}}
\newcommand{\EN}{\mbox{\tiny{EN}}}
\newcommand{\END}{\mbox{\tiny{EN}}}

\def \v{\mbox{var}}
\def \E{\mbox{E}}

\newcommand{\bomega}{\mbox{\boldmath $\bomega$}}

\newcommand{\bc}{\begin{center}}
\newcommand{\ec}{\end{center}}

\newcommand{\bi}{\begin{itemize}}
\newcommand{\ei}{\end{itemize}}

\newcommand{\be}{\begin{enumerate}}
\newcommand{\ee}{\end{enumerate}}

\newcommand{\bs}{\begin{slide*}}
\newcommand{\es}{\end{slide*}}

\newcommand{\bq}{ \begin{equation} }
\newcommand{\eq}{\end{equation}}

\makeindex

\begin{document}


\title{Handbook of Infectious Disease Data Analysis} 
\author{Jon Wakefield, Tracy Qi Dong and Vladimir N. Minin}
\maketitle

\cleardoublepage

\chapter{Spatio-Temporal Analysis of Surveillance Data}
\large{\bf Abstract}

\normalsize
\noindent 
In this chapter we consider space-time analysis of surveillance count data. Such data are ubiquitous and a number of approaches have been proposed for their analysis. We first describe the aims of a surveillance endeavor, before reviewing and critiquing a number of common models. We focus on models in which time is discretized to the time scale of the latent and infectious periods of the disease under study. In particular, we focus on the time series SIR (TSIR) models originally described in \cite{finkenstadt:grenfell:00} and the epidemic/endemic models first proposed in \cite{held:etal:05}. We implement both of these models  in the {\tt Stan} software, and  illustrate their performance via analyses of measles data collected over a 2-year period in 17 regions in the Weser-Ems region of Lower Saxony, Germany.

\section{Introduction}

The surveillance of disease statistics is now routinely carried out at both the national and subnational level. For example, in the United States, the Centers for Disease Control and Prevention (CDC) has a national notifiable disease surveillance system (NNDSS) to which case notifications for more than 70 infectious disease are sent  by all states. At a more local level, surveillance systems are also implemented by state public health departments.

There are many uses for a surveillance system. Early detection of outbreaks is clearly important, in order to quickly assign resources to minimize the disease burden and hopefully determine the cause(s) of the outbreak. Various approaches, with varying degrees of sophistication, are available. Often, no formal statistical methods are used, but rather astute public health workers notice increased counts. A simple approach is to analyze each area separately and to compare newly collected data with the historic numbers of cases \cite{levin:etal:15}. A more refined approach  is to use a scan statistic. For example, Greene et al \cite{greene:etal:16}, describe how the New York City Department of Health and Mental Hygiene carry out automated daily spatio-temporal cluster detection using the SatScan software, and discuss the action taken in response to several outbreaks including three common bacterial caused infections that cause diarrhea: shigellosis, legionellosis and campylobacteriosis. This approach does not acknowledge that spread of an infectious disease under examination has complex nonlinear dynamics. The models we describe in this chapter have generally not been used for outbreak detection, but could be, if ``trained" on retrospective data.

Prediction of future disease counts will clearly be of interest in some situations and for this purpose a model that is not built around an infectious process may be adequate.
If one is interested in predictions under different scenarios, for example, following different vaccination strategies, then a biologically motivated model is likely to be more useful than a model without a strong link to the underlying science. The same is true of  another aim, which is to gain an understanding of disease dynamics, including estimation of fundamental parameters such as $R_0$.

In a time series only situation in which space is not considered, a number of authors have discussed integer-valued autoregressive (INAR) models in the context of modeling infectious disease data \cite{cardinal:etal:99,fokianos:fried:10,fernandez:etal:16}. The epidemic-endemic models \cite{held:etal:05}, that we discuss extensively in this chapter, are closely related to  integer-valued generalized autoregressive conditionally heteroscedastic (INGARCH) models, that have also been used for modeling infectious disease data \cite{ferland:etal:06,zhu:11}. Since these approaches do not consider spatial modeling we do not consider them further (though we note that there is no reason that they couldn't be extended to include a spatial component).




In a surveillance setting the following data are typically available: demographic information on each case, for example, age and gender;
symptom onset date; date of diagnosis; clinical information, for example, symptoms; laboratory information on {virology}, perhaps on a subset of cases; areal (ecological) geographical information. These data are usually supplemented with population information at the areal level. We emphasize that we will be concerned with the usual form in which the data are available, which is {\it incidence counts}, that is new cases of disease (as opposed to {\it prevalence data} which constitute the total counts).

We structure this chapter as follows. In Section \ref{sec:transmission}, we review disease transmission models, with  brief descriptions of deterministic approaches, discrete time and continuous time models. Section \ref{sec:motivating} introduces the measles data example, and Section \ref{sec:models} describes in detail discrete space-time models. We return to the measles data in Section \ref{sec:data} and conclude with a discussion in Section \ref{sec:discussion}. On-line materials contain code to reproduce all analyses.

\section{Overview of Disease Transmission Models}\label{sec:transmission}

\subsection{Deterministic Models}\label{sec:deterministic}

Historically \cite{kermack:mckendrick:27}, infectious disease data were analyzed using deterministic models based on differential equations, see \cite{anderson:may:91} for a thorough discussion. As an example, we consider the susceptible-infectious-recovered (SIR) model, which is depicted in Figure \ref{fig:SIR_flow}. Models are set up based on a set of compartments in which individuals undergo homogenous mixing. This approach is typically used when the number of disease counts is large, and the integer numbers in the constituent S, I and R compartments are taken to be continuous.
Let $x(t)$, $y(t)$, $z(t)$ be the number of {susceptibles}, {infectives}, {recovered} at time $t$ in a {closed population}. 
The {\it hazard rate (force of infection)} is
 $$\underbrace{\lambda^\dagger(t)}_{{\text{Hazard}}} = \underbrace{c(N)}_{{\text{Contact Rate}}} \times \underbrace{
 \frac{y(t)}{N}}_{{\text{Prevalence}}} \times \underbrace{p_{{I}}}_{{\text{Infection Prob}}}.$$
Two common forms for the contact rate \cite{begon:etal:02} are
$$
c(N)=
\left\{ 
 \begin{array}{lr}
c_{\mbox{\tiny{FD}}} & \mbox{Frequency Dependent},\\
N c_{\mbox{\tiny{DD}}} & \mbox{Density Dependent}.
\end{array}
\right.
$$
The frequency-dependent model is often used, particularly for childhood infections when the most relevant contact group is the classroom, whose size will be of the same order, regardless of the population size. Under frequency dependency, $\lambda^\dagger(t) = \beta y(t)/N,$
where $\beta = c_{\mbox{\tiny{FD}}}  \times p_\text{I}$.

The deterministic SIR model is defined through classic {\it mass-action} \cite{anderson:may:91}. With frequency dependent transmission we have the following set of ordinary different equations:
\begin{eqnarray*}
\frac{dx(t)}{dt} &=& -\frac{\beta x(t) y(t)}{N}, \\
\frac{dy(t)}{dt} &=& \frac{\beta x(t) y(t)}{N} - \gamma y(t),\\
\frac{dz(t)}{dt} &=& \gamma y(t) ,
\end{eqnarray*}
with {\it  infection rate} $\beta$ and {\it recovery rate} $\gamma$. 

\begin{figure}[h!]
\centering
\begin{tikzpicture}[ edge from parent/.style={draw,-latex}, level/.style={sibling distance=25mm/#1}, node distance=1.5cm]
 \tikzstyle{v}=[rectangle, draw, thick, minimum height=1.25cm, minimum width=1.25cm]
   \node[v] (Su)  {$x(t)$};
   \node[scale=1] at (0,1.5) {{\bf S}};
   \node[circle, right of=Su] (i) {};
   \node[v, right of=i] (I) {$y(t)$}; 
            \node[scale=1] at (3,1.5) {{\bf I}};
   \node[circle, right of=I] (r) {};
   \node[v, right of=r] (R) {$z(t)$}; 
   \draw[thick,->]
        (Su) to node[above,scale=.8] {$\beta x(t)y(t) /N$ } (I);
    \draw[thick,->]
        (I) to node[above,scale=.8] {$\gamma y(t)$} (R);
              \node[scale=1] at (6,1.5) {{\bf R}};
\end{tikzpicture}
\vspace{.4in}
\caption{Susceptible-Infectious-Recovered (SIR) model representation. Solid arrows show the movement from S to I to R.
}\label{fig:SIR_flow}
\end{figure}
To turn these equations into a statistical model there are two important considerations:
\begin{enumerate}
\item Given initial states, and values for the parameters $\beta$ and $\gamma$ these differential equations can be solved to find the time trajectories of the three compartments. This must be done numerically, but can be achieved very efficiently, which means, in general, that complex compartmental models can be formulated, with the advantage that the transmission parameters are biologically interpretable.
\item More critically, the introduction of an artificial error model is needed. For example, one could assume additive errors with constant variance or a variance that depends on the mean, and then fitting can be performed in a straightforward fashion using ordinary or weighted least squares. The arbitrariness of the error model means that inference is dicey. Implicitly considering the inherent stochasticity directly is important for small populations and when the disease is rare. Ordinary or weighted least squares can be used for fitting, with the implicit assumption of uncorrelated errors with constant  variance, see for example \cite{ciupe:etal:06}, or refined versions of least squares \cite{hooker:etal:11}. 
The  arbitrariness of the (implicit) error model is unlikely to result in reasonable uncertainty quantification for parameter estimation.
For these reasons, we do not consider these models further here.
\end{enumerate}

\subsection{Discrete-Time Stochastic Models} 

We will concentrate on discrete-time models, and postpone an in-depth discussion of these models to Section \ref{sec:models}. The basic idea is to model the current disease counts as a function of previous counts, on a regular time scale.
For a discrete-time stochastic SIR model  one may choose the time scale to equal the transmission dynamics scale (latency plus infectious periods) or the generation time (time from infection of a primary case to infection of a secondary case infected by the primary case) \cite{nishiura:etal:10}. For example, often, but not always, 2 weeks is used for measles. Let $X_t$ and $Y_t$ be random variables representing the number of susceptibles and infectives at time $t$, $t=1,\dots,T$.  In the simplest case, the counts at $t$  depend only on the counts at the previous time $t$. Susceptibles may be reconstructed from,
 ${X_t} = {X_{t-1}} - {Y_t},$
 assuming a closed population (we describe more complex susceptible reconstruction models in Section \ref{sec:models}). The joint distribution of the counts is,
$$\Pr(y_1,\dots,y_T,x_1,\dots,x_T | y_0) = \prod_{t=1}^T \Pr(y_t | y_{t-1} ,x_{t-1}) \times \Pr(x_t | y_t, x_{t-1}),$$
where the second term is deterministic and we have suppressed the dependence of the first term on unknown parameters.
 
\subsection{Continuous-Time Stochastic Models}

The most realistic approach is to build a model that considers infections and recoveries on a continuous time scale. 
We describe a  continuous-time Markov chain for $\{X(t),Y(t), t \geq 0\}$ with frequency dependent transmission.
The  {\it transition probabilities} for a
susceptible becoming infective  and an infective  becoming recovered are:
\begin{eqnarray*}
\Pr\left(~ \begin{bmatrix} X(t+\Delta t)\\Y(t+\Delta t) \end{bmatrix} = \begin{bmatrix}{x-1}\\{y+1}
\end{bmatrix} ~\left|~ \begin{bmatrix} X(t)\\ Y(t) \end{bmatrix} = \begin{bmatrix}{x}\\{y}\end{bmatrix} \right.~\right) &=&\frac{\beta  {x}{y}}{N}\Delta t + o(\Delta t),\\
\Pr\left(~ \begin{bmatrix} X(t+\Delta t)\\Y(t+\Delta t) \end{bmatrix} = \begin{bmatrix}{x}\\{y-1}
\end{bmatrix} ~\left|~ \begin{bmatrix} X(t)\\ Y(t) \end{bmatrix} = \begin{bmatrix}{x}\\{y}\end{bmatrix} \right.~\right) &=&\gamma{y}  \Delta t + o(\Delta t),\\
\end{eqnarray*}
where the remainder terms $o(\Delta t)$ satisfy $o(\Delta t)/\Delta t \rightarrow 0$ as $\Delta t \rightarrow 0$. From the standpoint of an infective, each can infect susceptibles in $\Delta t$ with rate $\beta \Delta t x/N$, where $x$ is the number of susceptibles at time $t$. From the standpoint of a susceptible, each can be infected in $\Delta t$ with rate $\beta \Delta t y/N$, where $y$ is the number of infectives at time $t$. This set-up leads to exponential times in each of the S and I compartments.  Interpretable parameters are contained in this formulation, but unfortunately this approach is not extensively used as it quickly gets computationally hideous as the populations increase in size, which is the usual case with surveillance data, see the references in \cite{fintzi:etal:17}. We do not consider these models further here.

\section{Motivating Data}\label{sec:motivating}

\begin{figure}[htbp]
	\centering
	\includegraphics[scale=0.09]{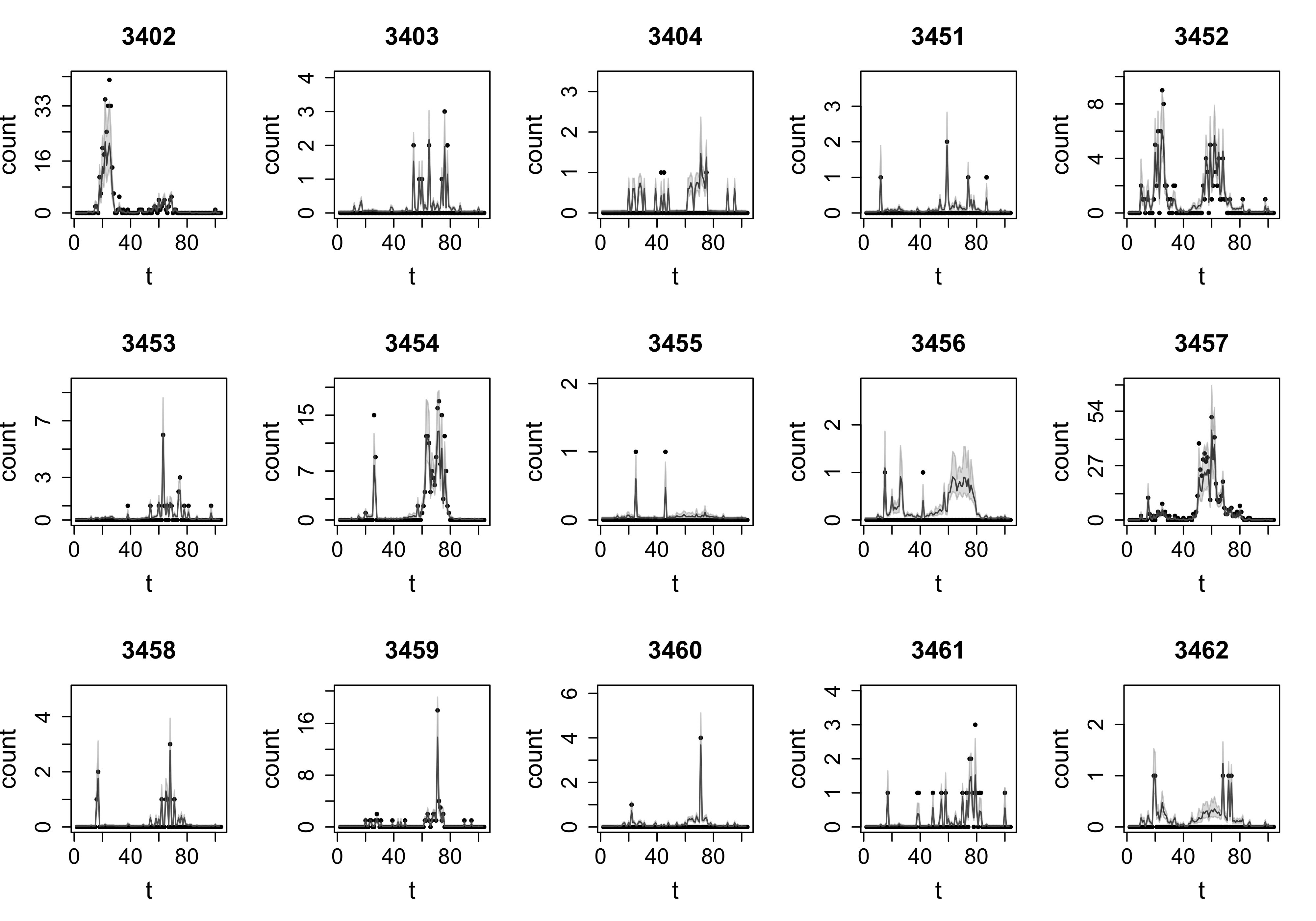}
	\caption{Observed (black dots) data in the 15 districts with non-zero counts, and posterior summaries (2.5\%, 50\%, 97.5\% quantiles) for $\mu_{it}$, under the TSIR model.}
	\label{fig:observed_fitted}
\end{figure}

\begin{figure}[htbp]
	\centering
	\includegraphics[scale=0.12]{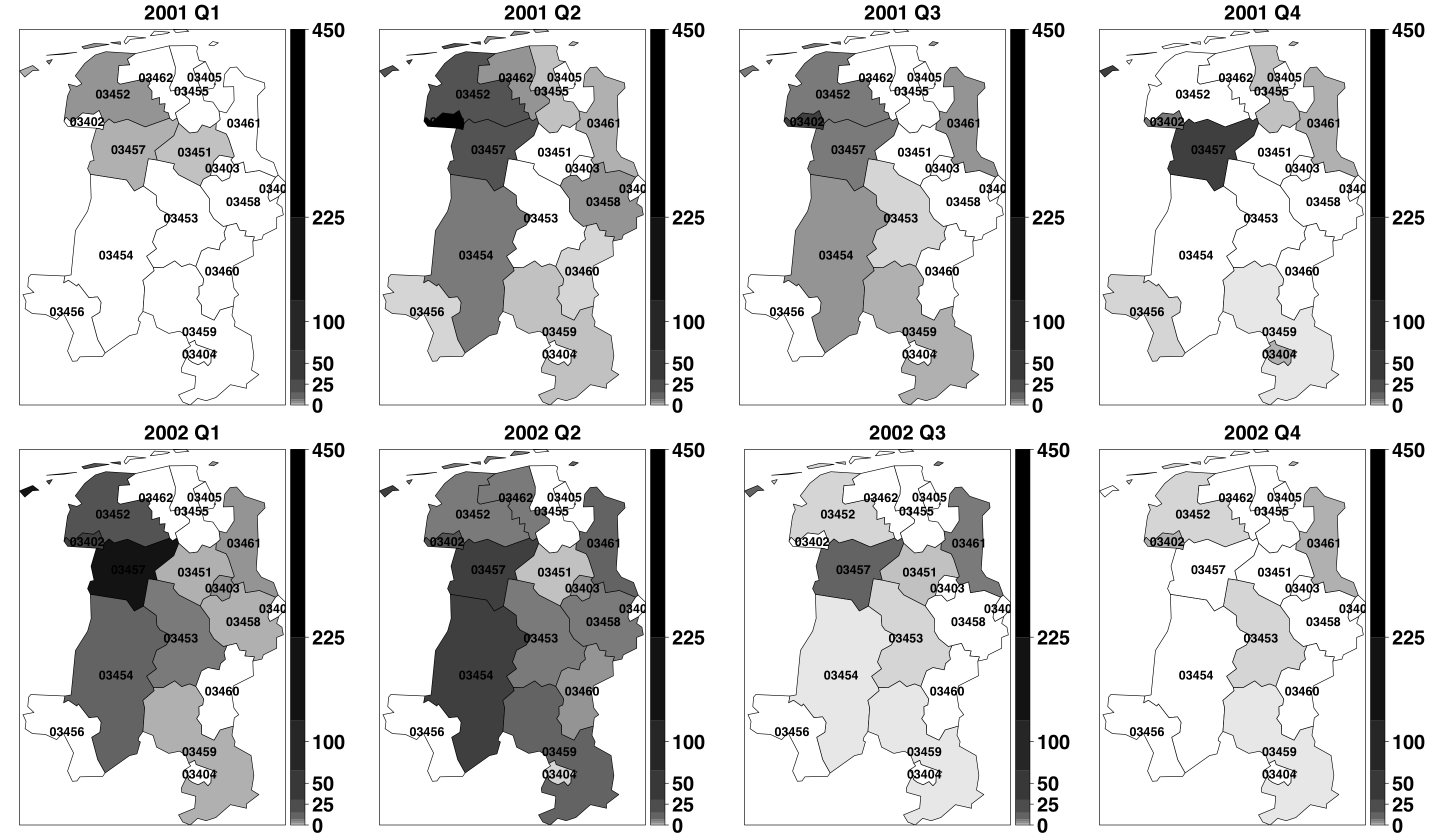}
	\caption{Average quarterly incidence by city district (per 100,000 inhabitants).}
	\label{fig:initial-maps}
\end{figure}

We analyze a simple measles dataset that has been extensively analyzed using the {\tt hhh4} framework that we describe in Section \ref{sec:hhh4}. The analysis of these data is purely illustrative and for simplicity we do not use the vaccination information that is available with these data (though we do discuss how such data may be included in Section \ref{sec:ecological}). We also analyze the data at a weekly time scale, for consistency with previous analyses of these data using epidemic/endemic models \cite{held:etal:05,meyer:etal:17}.
There is no demographic information on the cases, no information on births; looking at measles post-vaccination in the developed world is unlikely to yield any insight into transmission.
Figure \ref{fig:observed_fitted} shows 15 time series of counts in 15 districts (we exclude 2 areas that have zero counts), and Figure \ref{fig:initial-maps} shows maps of total cases over 3-month intervals. We see great variability in the numbers of cases over time and area. 

\section{Discrete-Time Spatial Models}\label{sec:models}

\subsection{Preliminaries}

We will derive  the probability that a susceptible individual at time $t-1$ will become infected by time $t$.
We assume that infected individuals are infectious for one time unit, before becoming removed, so that we have an {SIR model} with a fixed infectious period duration. Hence,
we lose the recovery rate parameter and incidence is assumed equal to prevalence.
All of the models that we discuss in this section assume that the force of infection is constant over the chosen time interval. A simple discrete-time model for the susceptibles in the context of measles \cite{finkenstadt:grenfell:00} is,
\begin{equation}\label{eq:susc}
X_t = X_{t-1} - Y_t + B_{t-d},
\end{equation}
where $B_{t-d}$ is the number of births $d$ time units previously, with $d$ chosen to be the number of time units for which maternally derived immunity lasts. There is no term for deaths since measles is primarily  a childhood disease and deaths from measles {are} low in the developed world setting in which our data are collected, as is child mortality.

\subsection{TSIR Models}

The time series SIR (TSIR) model was first described in \cite{finkenstadt:grenfell:00}, and has received considerable attention. 
The TSIR framework usually uses a negative binomial model, which we now carefully define, since there is some confusing terminology.  To quote from \cite{bjornstad:etal:02} (with our own italics added), ``Starting with $I_t$ infected
individuals, and assuming independence between
them, the {\it birth-death process} will hence be realized
according to a negative binomial distribution." In the following, we will consider a simple birth process. Perhaps the death process that is mentioned in this quote is referring to the infectives becoming recovered after one time unit. As \cite{kendall:49} comments (page 236), ``In the deterministic theory it made no difference whether the intrinsic rate of growth $\nu$ was purely reproductive in origin, or was really a balance, $\beta - \mu$, between a birth rate $\beta$ and a death rate $\mu$. In the stochastic theory this is no longer true, and the birth-and-death process is quite distinct from the pure birth process just described".

A negative binomial model is developed for the number of infectives and so we start with a review of this distribution. In one parameterization of the negative binomial model, the constant $r$ is the number of failures until an experiment is stopped and $k=0,1,\dots$ is the number of successes that occur on the way to these $r$ failures, with $p$ the success probability. It turns out that this scenario does not align with the present context, but it is the standard motivation, so we keep this language for a short while.
The negative binomial probability mass function is defined by
\begin{equation}\label{eq:negbin0}
 \Pr(K=k) = \binom{k+r-1}{k} (1-p)^rp^k.
\end{equation}
We write as $K \sim \mbox{NegBin}(\mu, r)$, and state the mean and variance,
\begin{eqnarray*}
\mbox{E}(K) &=& \frac{rp}{1-p} =\mu,\\
\mbox{Var}(K) &=&  \frac{rp}{(1-p)^2} = \mu \left( 1+ \frac{\mu}{r} \right).
\end{eqnarray*}
Note that $p=\mu/(\mu+r)$. The overdispersion relative to the Poisson is indexed by $r$, with greater overdispersion corresponding to smaller $r$. The negative binomial can be formulated in a number of ways; the total number of trials $K^\star=K+r$, may also be described as negatively binomially distributed with $K^\star=r,r+1,r+2,\dots$. We write, $K^\star \sim  \mbox{NegBin}^\star(\mu^\star, r)$,  where $\mu^\star = r+\mu = r/(1-p)$.

Moving towards the context of interest,
the negative binomial distribution arises as the distribution of the population size in a linear birth process, see for example \nocite{feller:50}Feller (1950, p.~448) and \nocite{cox:miller:65}Cox and Miller (1965, p.~157).
 We describe a linear birth process, also known as a Yule-Furry process.
 Individuals currently in the population each reproduce independently in $(t,t+\Delta t)$ with probability $\alpha \Delta t+o(\Delta t)$.
 Let $N(t)$ be the number of individuals at time $t$.
 Then,
$$\Pr(\mbox{ birth in }(t,t+\Delta t) ~| ~N(t)=n~) = n\alpha \Delta t + o (\Delta t),$$
with the probability of two or more births being $o(\Delta t)$.
 Let,
$$\Pr( N(t)= n \mid N(0) = n_0) =p_n(t).$$
 From the Kolmogorov forward equations,
\begin{eqnarray*}
p_n(t+\Delta t) = p_n(t) (1-n\alpha \Delta t) + p_{n-1}(t)(n-1) \alpha \Delta t + o(\Delta t),
\end{eqnarray*}
where the $n-1$ appears in the second term on the RHS because any of the $n-1$ individuals could give birth and each does so with probability $\alpha$.
Hence,
$$p'_n(t) = -n \alpha p_n(t) +(n-1) \alpha  p_{n-1}(t),
$$
with $p_n(0) = 1$ for $n=n_0$ and $p_n(0) = 0$ for $n \neq n_0$.
It can be then verified directly (Feller) or using probability generating functions (Cox and Miller) that for $n \geq n_0 > 0$:
\begin{equation}\label{eq:negbin}
p_n(t) = \Pr( N(t) = n) =\binom{n-1}{n-n_0} \mbox{e}^{- \alpha t n_0}\left(1-\mbox{e}^{- \alpha t}\right)^{n-n_0},
\end{equation}
with the probability $p$ in (\ref{eq:negbin0}) given by $p_t=1-\mbox{e}^{-\alpha t }$ and $N(t) \sim \mbox{NegBin}^\star( \mu^\star_t , n_0~)$ with $\mu_t^\star = n_0 \mbox{e}^{\alpha t}$.
This is also derived for $n_0=1$ by \nocite{kendall:49}Kendall  (1949, equation (17)).
  Note that the total number of trials $N(t)$ corresponds to $r+K$, which fits in with the total population size $n$ (so there are $n-n_0$ births).
We can also think of this model as starting with $n_0$ individuals, each of which independently produces a geometric number of progenies by time $t$; the probabilities are given by (\ref{eq:negbin}) with $n_0=1$.  

Now let $M(t)=N(t)-n_0$ be the number of births since time 0. Then,
\begin{equation}\label{eq:negbin-births}
\Pr( M(t) = n) =\binom{n + n_0 -1}{n} \mbox{e}^{- \alpha t n_0}\left(1-\mbox{e}^{- \alpha t}\right)^{n},
\end{equation} 
so $M(t) \sim \text{NegBin}\left(\mu_t,  n_0\right)$, with  
$$\mu_t = \mbox{E}(M(t) ) =\frac{n_0 p_t}{1-p_t} = n_0 (\mbox{e}^{\alpha t}-1),$$
and 
$$ \mbox{Var}(M(t) )= \frac{n_0 p_t}{(1-p_t)^2} =n_0(\mbox{e}^{2\alpha t}-\mbox{e}^{\alpha t})
= \mu_t\left(1+\frac{\mu_t}{n_0}\right)
,$$
so that as $n_0 \rightarrow \infty $ we approach the Poisson distribution.

In the context of the SIR model, let $y_{t-1}$ be the number of infectives at time $t-1$ and then assume that each gives rise to infectives with constant rate $x_{t-1}\beta/N$ over the interval $[t-1,t)$ (where we have assumed frequency dependent transmission). Note that the constant hazard is an approximation, because as a new infective appears in $[t-1,t)$ the number of susceptibles $x_{t-1}$ will drop by 1.
 With respect to the linear birth process derivation, $y_{t-1}$ is the ``initial number".

We assume that the previous infecteds are all recovered.
Then the total number of infectives is equal to the number of new infectives  that are produced $Y_t$ 
is a negative binomial random variable, NegBin$(\mu_t,y_{t-1})$, with
$$
 \Pr( Y_{t} = y_t \mid Y_{t-1} = y_{t-1}  ) =\binom{y_t+y_{t-1}-1}{y_{t}}\left( \mbox{e}^{- \beta x_{t-1}  /N}\right)^{y_{t-1}}\left(1-\mbox{e}^{- \beta x_{t-1} /N}\right)^{y_t}
$$
and
 \begin{eqnarray*}
\mbox{E}(Y_t \mid Y_{t-1}  = y_{{t-1}}) &=& \mu_t  = y_{t-1} \left(\mbox{e}^{x_{t-1}\beta/N} - 1\right), \\
 \mbox{Var}(Y_t \mid Y_{t-1}  = y_{t-1}) &=&  \mu_t(1+\mu_t/y_{t-1}).
 \end{eqnarray*}
 The latter form is a little strange since the level of overdispersion is not estimated from the data.
 Note that as $y_{t-1} \rightarrow \infty$, the negative binomial tends to a Poisson.
 If $x_{t-1}\beta/N$ is small,
 $$\mbox{E}(Y_t \mid Y_{t-1}  {= y}_{{t-1}}) \approx y_{t-1} x_{t-1}\beta/N.
 $$



We now explicitly examine TSIR models, an umbrella term which includes a number of variants.
First, note that the susceptibles are often modeled as (\ref{eq:susc}), with correction for underreporting.
We begin with a single area and the form,
\begin{eqnarray*}
Y_{t} \mid Y_{t-1}  {= y}_{{t-1}}, X_{t-1}  {= x}_{{t-1}} \sim \mbox{NegBin}(  \mu_{t}, y_{t-1} ),
\end{eqnarray*}
where
$$
\mu_t = \beta y_{t-1}^\alpha x_{t-1}^\gamma/N.$$
The power $\gamma$ is far less influential than the power $\alpha$ \cite{liu:etal:87} and so in general the case $\gamma=1$ has been considered, with $\alpha$ set to be just below 1 (which slows down the spread).  The rationale for the power $\alpha$ is that it is  included to allow for deviations from mass action \cite{finkenstadt:grenfell:00} and to account for the discrete-time approximation to the continuous time model \cite{glass:etal:03}.  With the non-linear term $y_{t-1}^\alpha$, the pure birth process approximating the SIR model is no longer linear. 
As a result, the number of births during a finite time interval are no longer negative binomially distributed results, but the TSIR authors still use the negative binomial nonetheless.
From the perspective of the infectives we have a generalized birth process.
We stress again that the level of overdispersion is not estimated from the data, but is determined by {$y_{t-1}$}. 
%

We now turn to the case where we have incidence counts $y_{it}$ indexed by both time $t$ and area $i$, $i=1,\dots,n$. Consider the model, $Y_{it} | y_{i,t-1},x_{i,t-1} \sim \mbox{NegBin}(\mu_{it},y_{i,t-1})$, with
$$\mu_{it} = \beta (y_{i,t-1}+\iota_{i,t-1})^\alpha x_{i,t-1}/N_i,$$
where
$\iota_{i,t-1}$ are infected contacts from areas other than $i$ \cite{bjornstad:grenfell:08}. Again, the negative binomial distribution does not arise mechanistically when the hazard is non-linear.

In \cite{xia:etal:04}, a ``gravity model" is assumed with $\iota_{i,t-1} \sim \mbox{Gamma}( m_{i,t-1}, 1),$
a gamma distribution with mean and variance $$
m_{i,t-1} = \upsilon N_{i}^{\tau_1} \sum_{j \neq i} \frac{y_{j,t-1}^{\tau_2}}{d_{ij}^{\rho}},
$$
where $d_{ij}$ is the distance between areas $i$ and $j$, and $\rho>0$ determines the strength of the neighborhood flow, with the limit as $\rho \rightarrow 0$ giving equal weight to all neighbors. In practice, at least a subset of these parameters are often fixed. For example, we could take $\tau_1=\tau_2=1$. A simpler model (that we fit in Section \ref{sec:data}) with these constraints might replace $\iota_{i,t-1}$ by its expectation to give the mean model,
\begin{eqnarray}
\mu_{it} &=& \beta \left(y_{i,t-1}+\upsilon N_{i} \sum_{j \neq i} \frac{y_{j,t-1}}{d_{ij}^{\rho}}\right)^\alpha \frac{x_{i,t-1}}{N_i}\label{eq:tsir1},\\
 &=&  \left( \mbox{e}^{\lambda^{\AR}}  y_{i,t-1}+\mbox{e}^{\lambda^{\NE}} N_{i} \sum_{j \neq i} \frac{y_{j,t-1}}{d_{ij}^{\rho}}\right)^\alpha \frac{x_{i,t-1}}{N_i}. \label{eq:tsir2}
\end{eqnarray}
Note that if for some $t$, $y_{it}=0$, for all $i$, then $\mu_{is}=0$ for $s>t$.  To prevent this issue, one might add an endemic term, $\mbox{e}^{\lambda^{\END}}$, to the mean (\ref{eq:tsir2}). This difficulty has been circumnavigated by randomly simulating a new count from a Poisson distribution with a mean that depends on the under-reporting rate, see the supplementary materials of \cite{vanboeckel:etal:16}.

In Section \ref{sec:data}, we fit a TSIR like model which takes various components from the epidemic/endemic model. Specifically, we fit the model,
\begin{align*}
Y_{it} | \mu_{it} & \sim \mbox{NegBin} \left( \mu_{it}, \phi \right) \\
\mu_{it} &= \left[ \mbox{e}^{ \lambda^{\AR}_t} y_{i,t-1} + \mbox{e}^{ \lambda^{\NE}} N_i^{\tau_1} \sum_{j \ne i} w_{ij} y_{j, t-1}^{\tau_2} \right]^{\alpha} + \mbox{e}^{ \lambda^{\EN}} N_i 
\end{align*}
with seasonality included in the AR component:
$\lambda_t^{\AR}  = \beta^{\AR}_0 + \beta^{\AR}_1 t + \gamma \sin(\omega t) + \delta \cos(\omega t)$. 
We have, replaced $y_{i, t-1}$ by $\phi$ as the overdispersion parameter $\iota_{i, t-1}$ equal to its mean $m_{i, t-1}$, set $x_{i, t-1}$ equal to $N_i$ (that is, approximating the susceptibles by the population), added an endemic term to the model, use normalized weights: 
$$w_{ij} = \frac{d_{ij}^{-\rho_1}}{\sum_{k \neq i} d_{ik}^{-\rho_1}},$$ 
and reparameterized the decay parameter as $\rho_1 = \theta_1/(1-\theta_1)$, with $0<\theta_1<1$.

A variety of fitting procedures have been used, with various levels of sophistication.
An MCMC scheme was used for a TSIR model  with under-reporting and an endemic term (called an influx parameter) appearing in the mean function \cite{morton:finkenstadt:05}. The under-reporting was accounted for by using an auxiliary variable scheme to impute the unknown true counts. This scheme is natural from a Bayesian persepctive, though computationally prohibitive in large populations.







TSIR models have been used to model data on multiple strains;  for dengue without a spatial component \cite{reich:etal:13} and with  spatial component for hand, foot and mouth disease \cite{takahashi:etal:16}. In the latter, the effect of vaccination was examined; the statistical analysis was based on estimated counts for EV71 and CoxA16 pathogens. A standard TSIR area model was used with no neighborhood component.



TSIR models have been used to examine various aspects of disease dynamics.
Koelle and Pascual \cite{koelle:pascual:04} use a nonlinear time series model to reconstruct patterns of immunity for cholera and examine the contributions of extrinsic factors such as seasonality (there are no spatial effects in the model). 
Seasonality of transmission has been considered by a number of authors, in particular in the context of measles. 
Climatic conditions that are more or less favorable to transmission might be relevant, along with other seasonal factors that might affect the extent of social contacts.
In contrast with the epidemic/endemic model, seasonality has been modeled in the autoregressive component in the TSIR framework. 
Seasonal forcing (the increase in transmission when children aggregate in schools, with a decrease during school holidays) is a particular aspect that has been considered.  A local smoothing model has been used on the raw rates \cite{london:yorke:73}.
The simple sinusoidal model  has been argued against \cite{earn:etal:00} as being too simplistic. 
The rate $\beta_t$ has been modeled as a function of month for rubella in Mexico \cite{metcalf:etal:11rubellamexico}, or allowing a unique set of 26 parameters in each year \cite{bjornstad:etal:02}.











 The TSIR model may also be fitted in {\tt R} in the {\tt tsiR} package \cite{becker2017tsir}, with implementation based on the method described in \cite{finkenstadt:grenfell:00} or a Bayesian implementation (using the {\tt rjags} package). At the time of writing, the models are  temporal only, with no explicit spatial modeling possible (other than fitting separate TSIR models in each area).

\subsection{Epidemic/Endemic {\tt hhh4} Models}\label{sec:hhh4}

We now describe a parallel development, originating in \cite{held:etal:05}, with subsequent developments being reported in \cite{paul:etal:08,paul:held:11,held:paul:12,meyer:held:14,geilhufe2014power,meyer:held:17}. For an excellent review of the statistical aspects of infectious disease modeling, and this class of models in particular, see \cite{hohle:16}.
The epidemic/endemic description is used to denote the addition of a term in the mean function that does not depend on previous counts, while {\tt hhh4} is the key function in the {\tt surveillance} package that fits the models we describe in this section.

We will derive  the probability that a susceptible individual at time $t-1$ will become infected by time $t$.
In contrast to the TSIR derivation, in which the process of the infectives infecting susceptibles was modeled (and lead to a negative binomial for the number of new infectives), the derivation here models the process of susceptibles becoming infected (and, as we will see, leads to a binomial distribution for the number of new infectives).

We again  assume that infected individuals are infectious for one time unit, before becoming removed, so that we have an {SIR model} with a fixed infectious period duration (and a constant hazard). So the event of a susceptible at $t-1$ becoming infected in  $[t-1,t)$ is  Bernoulli with probability of infection,
$1-\exp(-\beta y_{t-1}/N)$. 




Under homogenous mixing and independence of the Bernoulli outcomes of the susceptibles,
\begin{equation}\label{eq:binomial}
Y_t | y_{t-1},x_{t-1} \sim \mbox{Binomial}\left( x_{t-1} ,{1-\exp(-\beta y_{t-1}/N)} \right).
\end{equation}
If we write $\eta = \exp (-\beta/N)$ (which is appropriate given the frequency dependent model we described in Section \ref{sec:deterministic}), we see we have a (Reed-Frost) chain binomial model (e.g.,~Daley and Gani, 1999, Chapter 4\nocite{daley:gani:99}).  Under the chain-binomial formulation a susceptible at time $t-1$ can remain susceptible by avoiding being infected by all infectives $y_{t-1}$, and the probability of avoiding being infected by one infective is $\eta$.
This leads to
$
Y_t | y_{t-1},x_{t-1} \sim \mbox{Binomial}( x_{t-1} ,1-\eta^ {y_{t-1}})
$, i.e.,
$$ \Pr( Y_t=y_t | y_{t-1},x_{t-1}) = 
\binom{x_{t-1}}{x_{t-1}-y_{t}} (\eta^{y_{t-1}})^{x_{t-1}-y_{t}} (1-\eta^{y_{t-1}})^{y_t}.
$$
%

Under the assumption of a rare disease, and the approximation $\exp(-\beta y_{t-1}/N) \approx 1- \beta y_{t-1}/N$ the model
(\ref{eq:binomial}) becomes (suppressing the dependence on $x_{t-1}$),
$Y_t | y_{t-1} \sim \mbox{Poisson}({\beta x_{t-1}y_{t-1}}/N).$
A further assumption that is implicitly made is that $x_{t-1} \approx N$ to give,
$Y_t | y_{t-1} \sim \mbox{Poisson}(\beta y_{t-1}).$ Hence, in the epidemic/endemic development there is an implicit assumption of frequency dependent transmission.

In the original paper \cite{held:etal:05}  a branching process with immigration formulation was taken:
$$Y_t = Y^\star_t+Z_t,$$
with independent components,
\begin{eqnarray*}
Y^\star_t | y_{t-1} &\sim& \mbox{Poisson}({\beta y_{t-1}}),\\
Z_t &\sim &\mbox{Poisson}(\upsilon_t),
\end{eqnarray*}
which are labeled {\it epidemic} and {\it endemic}, respectively. The epidemic component corresponds to the branching process, and the endemic component to immigration.
To account for overdispersion in the infectives count, the model is,
$$Y_t | y_{t-1} \sim \mbox{NegBin}( \mu_t, \kappa),$$
with 
\begin{eqnarray*} 
\mbox{E}(Y_t | y_{t-1}) &=& \mu_t = \beta y_{t-1},\\
\mbox{Var}(Y_t | y_{t-1}) &=& \mu_t\left(1+\frac{\mu_t}{\kappa}\right),
\end{eqnarray*}
where $\kappa$ is estimated from the data.
Unlike the TSIR development, this distribution does not ``drop out" of a stochastic process formulation, but is made on pragmatic considerations.
Negative binomial distributions have also been used in other infectious disease developments.
For example, in \cite{lloyd:etal:05} it is assumed that individual level reproductive numbers are gamma distributed, with a Poisson distribution for the cases infected by each infective, to give a negative binomial distribution for the number of infectives generated.

The development of the TSIR model  was based directly on the number of infectives produced by the current infectives. In contrast, the epidemic/endemic development here determines the risk of each susceptible being infected.
The negative binomial distribution derived under the TSIR framework has countably infinite support, whereas the number of susceptibles is bounded, but given the rate drops with the latter, this approximation is unlikely to be problematic. The negative binomial distribution of the TSIR is a continuous time Markov chain (CTMC) approximation, while the epidemic/endemic in this section is a discrete time Markov chain (DTMC).

Now we consider the more usual situation in which  we have incident counts $y_{it}$ and populations $N_{it}$ in a set of areas indexed by $i=1,\dots,n$. Suppose that new infections can occur:
\begin{enumerate}
\item from {\it self-area} infectives;
\item from {\it neighboring-area} infectives;
\item from another source, which may be an environmental reservoir, or infectives from outside the study region.
\end{enumerate}
In the case of different  possibilities for  becoming infected, we can use the classic {\it competing risks} framework \cite{prentice:etal:78}, in which the hazard rates (forces of infection) are additive. We let $\lambda^{\mbox{\tiny{TOT}}}_{it}$ represent the overall hazard for a susceptible in area $i$ at time $t$, and write
$$\lambda^{\mbox{\tiny{TOT}}}_{it} =  \underbrace{\lambda^{\mbox{\tiny{AR}}}_{it}}_{\text{{Self-Area}}}  +   \underbrace{\lambda^{\mbox{\tiny{NE}}}_{it}}_{\text{{Neighboring-Area}}} +  \underbrace{\lambda^{\mbox{\tiny{EN}}}_{it} }_{\text{{Environmental}}}.
$$
Assuming $\lambda_{it}^{\mbox{\tiny{TOT}}}$ is small, the probability of infection in $[t-1,t)$, for a single susceptible
is, 
$
 1-\exp(-\lambda_{it}^{\mbox{\tiny{TOT}}}) \approx  \lambda^{\mbox{\tiny{TOT}}}_{it} .
$
%
Following detailed arguments in \cite{bauer:wakefield:17} (including again assuming that $x_{i,t-1} \approx N_{it}$) we  obtain the conditional mean,
 \begin{equation*}
 \mu_{it} = \underbrace{ 
  \lambda^{\AR}_{it}y_{i,t-1}}_{\text{Self-Area}}  
+
\underbrace{   \sum_{j=1 }^n\lambda^{\NE}_{it} w_{ij}y_{j,t-1} }_{\text{Neighboring Areas}}+\underbrace{N_{it}\lambda^{\EN}_{it}}_{{\text{Environmental}}},
\end{equation*}
where $w_{ij}$ are a set of weights that define the neighborhood structure.
The rates may {depend} on both space and time to allow covariate modeling, trends in time (including seasonality) and area-specific random effects, which may or may not have spatial structure.
In practice, sparsity of information will lead to simplifications, in particular, we describe the model for the Germany measles data. There are {only} 17 areas  which (along with the low counts) means a simple neighborhood model is considered.
Specifically, as in the majority of illustrations of the epidemic/endemic framework we assume the seasonality model,
 \begin{equation*}
 \mu_{it} =  \lambda^{\AR}_{i}y_{i,t-1} +
\lambda^{\NE} \sum_{j=1 }^n  w_{ij}y_{j,t-1} +N_i \lambda^{\EN}_{it},
\end{equation*}
with
\begin{eqnarray*}
\log  \lambda^{\AR}_{i} &=& \beta_0^{\AR} + b_i^{\AR},\\
\log  \lambda^{\EN}_{it} &= &\beta_0^{\EN} + \beta_1^{\EN} t+ \gamma \sin (\omega t) + \delta \cos (\omega t) + b_i^{\EN},
\end{eqnarray*}
where $\omega=2\pi/26$ for biweekly data. Note the contrast with the TSIR framwork, in which seasonality was modeled in the autoregressive term, though we note that seasonality can be incorporated in any of the three terms \cite{held:paul:12}, and this possibility is available in the {\tt surveillance} package.

Originally, the weights were binary corresponding to spatial contiguity. 
More recently \cite{meyer:held:14}, the weights are assumed to follow a {power law}, with
$$w_{ij}=\frac{m_{ij}^{-\rho_2}}{\sum_{k \neq i} m_{ik}^{-\rho_2}},$$ 
and where $m_{ij}$ is the number of areas that must be crossed when moving between areas $i$ and $j$, and $\rho_2 \ge 0$ is a {power} that may be estimated. The normalization ensures that $\sum_{k \neq i} w_{ik}=1$ for all rows of the weight matrix (infecteds are being {allocated} to neighbors). 
The limit {$\rho_2 \rightarrow \infty$} corresponds to first-order dependency,  and 
{$\rho_2 = 0$} gives equal weight to all areas. 
The power law allows ``contact" between areas that are a large distance apart since it is ``heavy-tailed".
Hence, by analogy with common spatial models we have two parameters, $ \lambda^{\NE}$ that determines the magnitude of the contribution from the neighbors and $\rho_2$ that determines the extent of the neighbor contributions. We parameterize as $\rho_2 = \theta_2/(1-\theta_2)$ with $0 < \theta_2 <1$.

Fitting via {maximum} likelihood/Bayes is relatively straightforward. The epidemic/endemic model with random effects is implemented using penalized quasi-likelihood (PQL) in the {\tt surveillance} package in {\tt R}.
The computational burden is not greatly impacted by the data size and so infectious disease data from large populations is not problematic. In Section \ref{sec:data} we describe a Bayesian approach, with implementation in {\tt Stan}.

A very similar model to the above has also been independently developed and fitted to data on hand, foot and mouth disease in China  \cite{wang:etal:11}. The model allowed for infections from not only the immediately previous time periods, but also for periods further back.
%

A more complex model in which there is not a fixed latency plus infectiousness period has been considered \cite{lekone:finkenstadt:06}. In the context of a susceptible-exposed-infectious-recovered (SEIR) model, they develop a discrete-time approximation to a stochastic SEIR model, in which there is a data augmentation step that imputes the number of infected over time. Given these numbers, the counts in each compartment are available, and the likelihood is a product of three binomials (one each for  infected counts, first symptom counts, removed counts).

We describe how the approach of \cite{lekone:finkenstadt:06} could be used for the SIR model. The key difference is that the fixed latency plus infectious period is relaxed and so it is no longer assumed that incidence is equal to prevalence. Consequently, we  continue to let $X_t$, $Y_t$, $Z_t$ represent the numbers of susceptibles, incident cases, and number who recover at time $t$, but now let $I_t$ be the prevalence. The model consists of the two distributions:
\begin{eqnarray}
Y_t | I_{t-1}=i_{t-1},X_{t-1}=x_{t-1} &\sim &\mbox{Binomial}\left(x_{t-1},1- \exp(\beta i_{t-1}/N) \right)\label{eq:bin1}\\
Z_t | I_{t-1}=i_{t-1} &\sim &\mbox{Binomial}\left(i_{t-1},1- \exp(\gamma) \right),\label{eq:bin2}
\end{eqnarray}
so that $\gamma$ is the recovery rate, with $1/\gamma$ the mean recovery time. In this model, the reproductive number is $R_0=\beta/\gamma$.
Further, 
\begin{equation}\label{eq:SIRinc}
I_t=I_{t-1}+Y_{t-1}-Z_{t-1},
\end{equation}
subject to the initial conditions $x_0=N$ (the population size, say) and $I_0=a$. An obvious way to carry out computation is to introduce the unknown counts $Z_t$ into the model as auxiliary variables, from which the series $I_t$ can be constructed from (\ref{eq:SIRinc}). With the ``full data" the likelihood is simply the product over the binomials in (\ref{eq:bin1})--(\ref{eq:bin2}).
This scheme will be computationally expensive for long time series with large numbers of cases.

\vspace{.2in}
\noindent
{\bf Summary:} Both the TSIR and the epidemic/endemic formulations are approximations. The TSIR formulation (without the $\alpha$ power) is in continuous time and assumes the number of susceptibles remains constant between observation points. The epidemic/endemic formulation assumes the number of infectives is constant between observation time points.

\subsection{Under-Reporting}

Under-reporting is a ubiquitous problem in infectious disease modeling; this topic is covered in detail in Chapter ??. We describe an approach to under-reporting that has been used along with TSIR modeling, particularly for measles, leaning heavily on \cite{finkenstadt:grenfell:00}. We let $C_t$ represent the reported cases and $Y_t$ the true number of cases. Then, again using $X_t$ as the number of susceptibles, we generalize (\ref{eq:susc}) to,
\begin{equation}\label{eq:susc2}
X_t = X_{t-1} - \rho_t C_t + B_{t-d} +U_t,
\end{equation}
where $1/\rho_t$ is the under-reporting rate at time $t$, and $u_t$ are uncorrelated errors with $\E(U_t)=0$, $\v(U_t)=\sigma_u^2$, that acknowledge the errors that may creep into the deterministic formulation.  We write 
$X_t=\overline{X}+ Z_t$, substitute in (\ref{eq:susc2}) and successively iterate to give,
$$Z_t = Z_0 -  \sum_{s=1}^t \rho_s C_s+ \sum_{s=1}^t B_{s-d} +  \sum_{s=1}^t U_s.$$
From this equation, we see that if under-reporting is present and not accounted for, the difference between cumulative births and observed cases will grow without bound. Let,
$$\widetilde{C}_t =  \sum_{s=1}^t C_s,\quad \widetilde{B}_t =  \sum_{s=1}^t B_{s-d},\quad \widetilde{U}_t =  \sum_{s=1}^t U_s, \quad R_t = \sum_{s=1}^t (\rho_s- \rho) C_s,$$
where $\rho=\E(\rho_t)$ and $R_t \approx 0$ under a constant reporting rate.
Then, to estimate $\rho$ it has become common practice in TSIR work to regress the cumulative births on the cumulative cases,
$$ \widetilde{B}_t = Z_t-Z_0 + \rho \widetilde{C}_t - \widetilde{U}_t,
$$
to estimate $\rho$ as the slope.
Note that the cumulative errors $\widetilde{U}_t$ do not have constant variance, since $\v(U_t)=\sigma_u^2$, so a weighted regression is more appropriate.

With an estimate $\widehat{\rho}$, one may estimate the true number of cases as $Y_t=\widehat{\rho} C_t$, and use these cases in a TSIR model. This approach is often followed, though there is no explicit acknowledgement of the additional (beyond sampling) variability in $Y_t$.

\subsection{Ecological Regression}\label{sec:ecological}

Ecological bias occurs when individual and aggregate level (with aggregation here over space) inference differ. The key to its understanding  is to take an individual-level model and  aggregate to the area-level \cite{wakefield:08}. We report on recent work \cite{fisher:wakefield:17}, that examines the implications of ecological inference for infectious disease data.
Let $Y_{itk}$ be disease indicator for a susceptible individual $k$ in area $i$ and week $t$ and $z_{itk}$ be an individual-level covariate, $k=1,\dots,N_i$. We present a simple example  in which there is an autoregressive term only, and also assume  a rare disease. In this case, the individual-level model is,
$$ Y_{itk}| y_{i,t-1,k}  \sim \mbox{Bernoulli}\left({\lambda^{\AR}_{itk} y_{i, t-1}/N_{i}}\right),$$
with
$ \lambda^{\AR}_{itk} = \exp(\alpha+\beta z_{itk}). $
The implied {aggregate hazard rate} for area $i$ and time $t$ is then \cite{fisher:wakefield:17},
$$\overline{\lambda}^{\AR}_{it} = \exp(\alpha) \int_{A_i}   \exp(\beta z) g_{it}(z) dz,  $$
where $A_i$ represents area $i$ and $g_{it}(z)$ is the {within-area distribution} of $z$ at time $t$. 

As a  simple example consider the binary covariate, $z_{itk}=0/1$ (for example, the vaccination status of individual $k$ in area $i$ at time $t$). 
The aggregate consistent model, for $Y_{it}=\sum_{k=1}^{N_i} Y_{itk}$, is
\begin{eqnarray*}
Y_{it} | \overline{\lambda}^{\AR}_{it} &\sim& \mbox{Poisson}( \overline{\lambda}^{\AR}_{it} y_{i,t-1}/N_i)\\
\overline{\lambda}^{\AR}_{it} &=& N_i \left[ (1-\overline{z}_{it}) \mbox{e}^{\alpha} +\overline{z}_{it} \mbox{e}^{\alpha+ \beta }\right].
\end{eqnarray*}
A naive model would assume $\overline{\lambda}^{\AR}_{it} = \exp(\alpha^\star +\beta^\star \overline{z}_{it})$, where $\overline{z}_{it}$ is the area average.
This naive model is very different from the aggregate consistent model.

\section{Analysis of Measles Data}\label{sec:data}

We now return to the measles data and, for illustration, fit a TSIR model and an epidemic/endemic  model. We use a Bayesian approach due to its flexibility, and since  the {\tt rstan} package allows these models to be fitted with relative ease.  The package implements MCMC using the no U-turns version \cite{hoffman:gelman:14} of Hamiltonian Monte Carlo \cite{neal:2011}. The code, and comparison of parameter estimates for the epidemic/endemic  model with the PQL versions in the {\tt surveillance} package, is included in the supplementary materials.

\subsection{TSIR Model}

The TSIR type model we fit is,
\begin{align*}
Y_{it} | \mu_{it} & \sim \mbox{NegBin} \left( \mu_{it}, \phi \right), \\
\mu_{it} &= \left[ \mbox{e}^{ \lambda^{\AR}_t} y_{i,t-1} + \mbox{e}^{ \lambda^{\NE}} N_i^{\tau_1}
 \sum_{j=1}^n w_{ij} y_{j, t-1}^{\tau_2} \right]^{\alpha} + N_i\mbox{e}^{ \lambda^{\EN}},
\end{align*}
where,
$$
w_{ij} = \frac{d_{ij}^{-\theta_1/(1-\theta_1)}}{\sum_{k \neq i} d_{ik}^{-\theta_1/(1-\theta_1)}},$$
with $d_{ij}$ being the distance between areas $i$ and $j$, and
$\lambda^{\AR}_t  = \beta^{\AR}_0 + \beta^{\AR}_1 t + \gamma \sin(\omega t) + \delta \cos(\omega t)$. 
Relatively uninformative priors were used (zero mean normals with standard deviations of 10). The deviation from linearity parameter, $\alpha$, was restricted to lie in $[0.95,1]$ and was given a uniform prior on this support. 

\begin{figure}[htbp]
	\centering
	\includegraphics[scale=0.09]{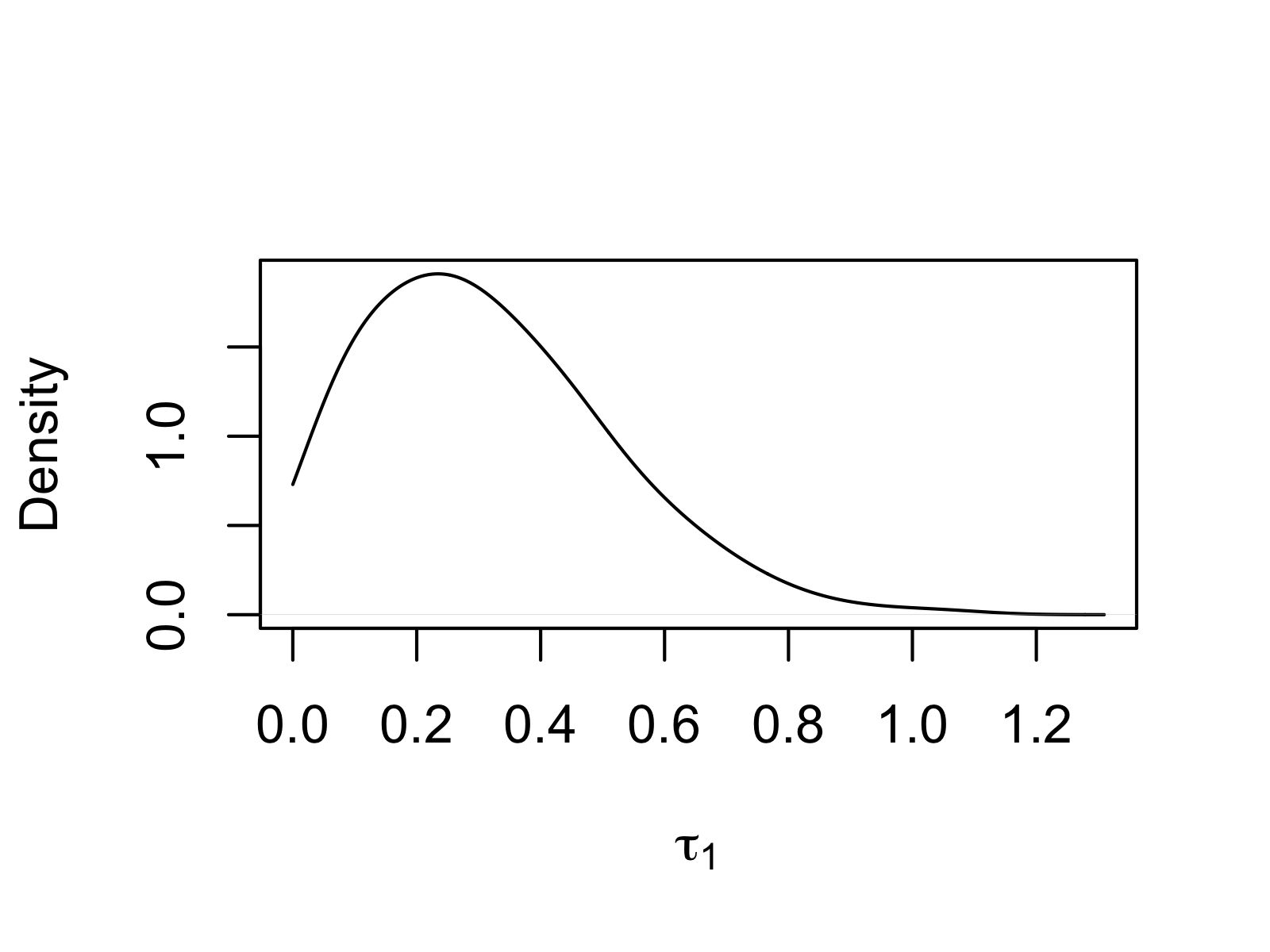}
	\includegraphics[scale=0.09]{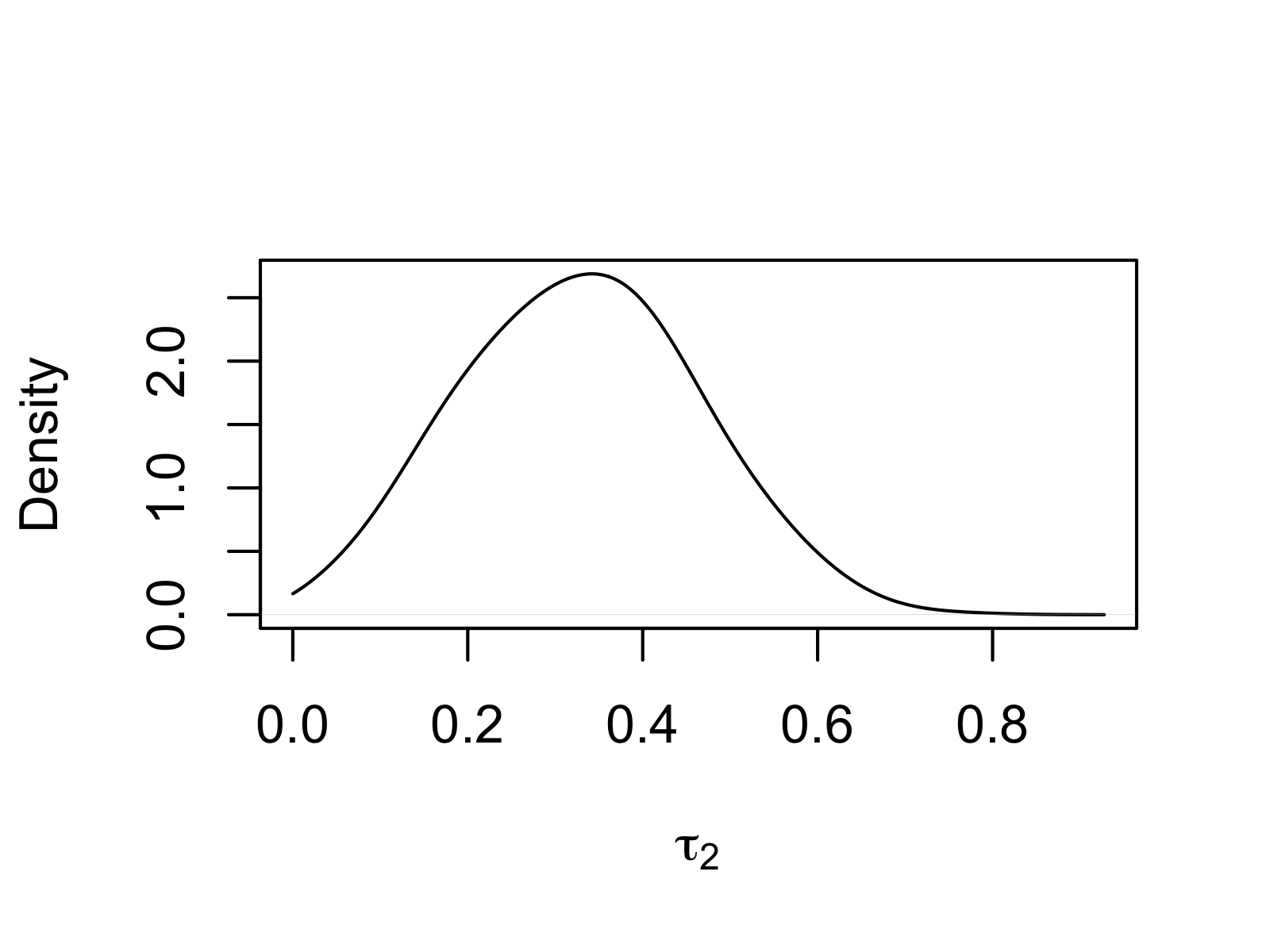}
	\includegraphics[scale=0.09]{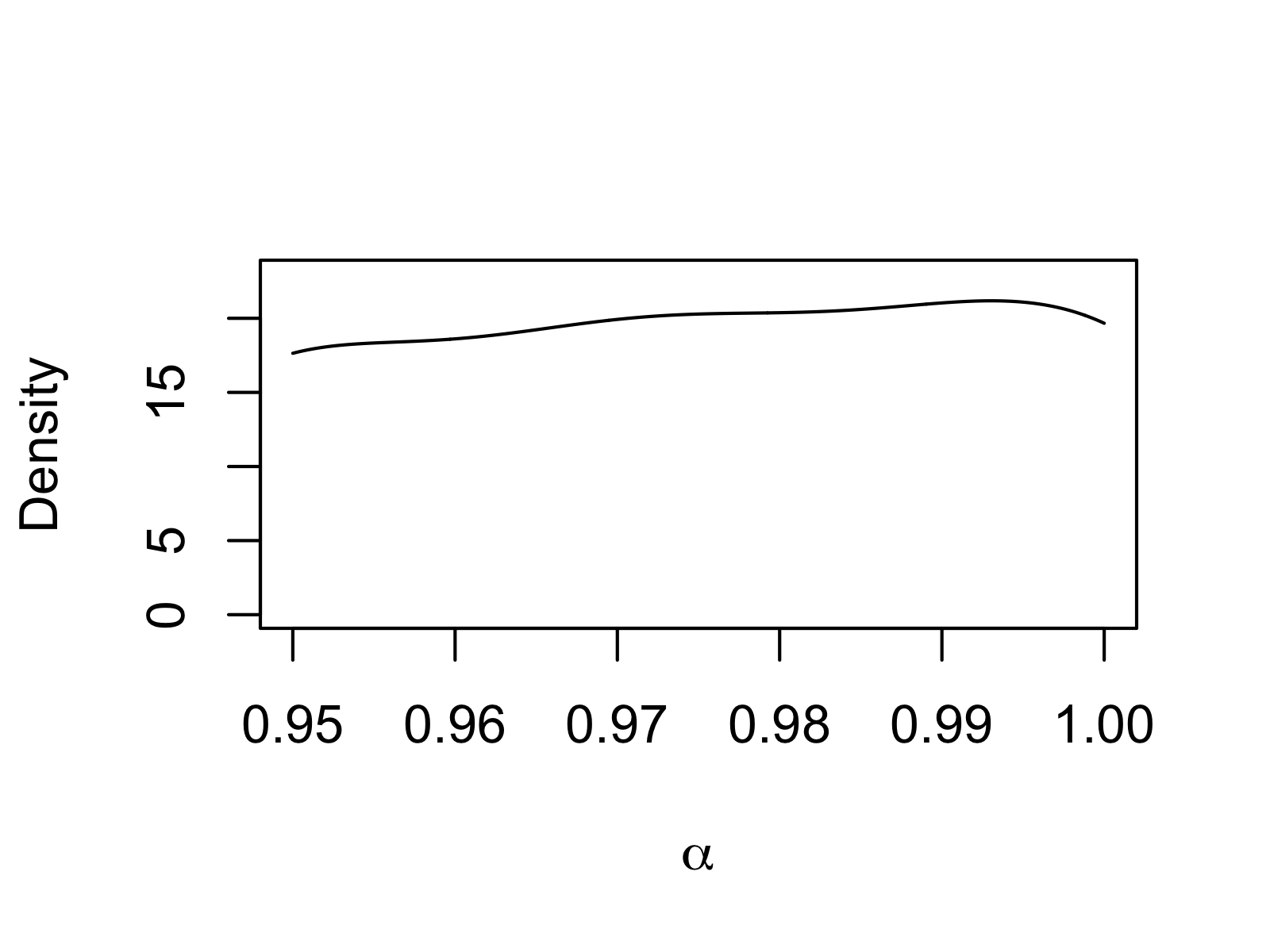}
	\includegraphics[scale=0.09]{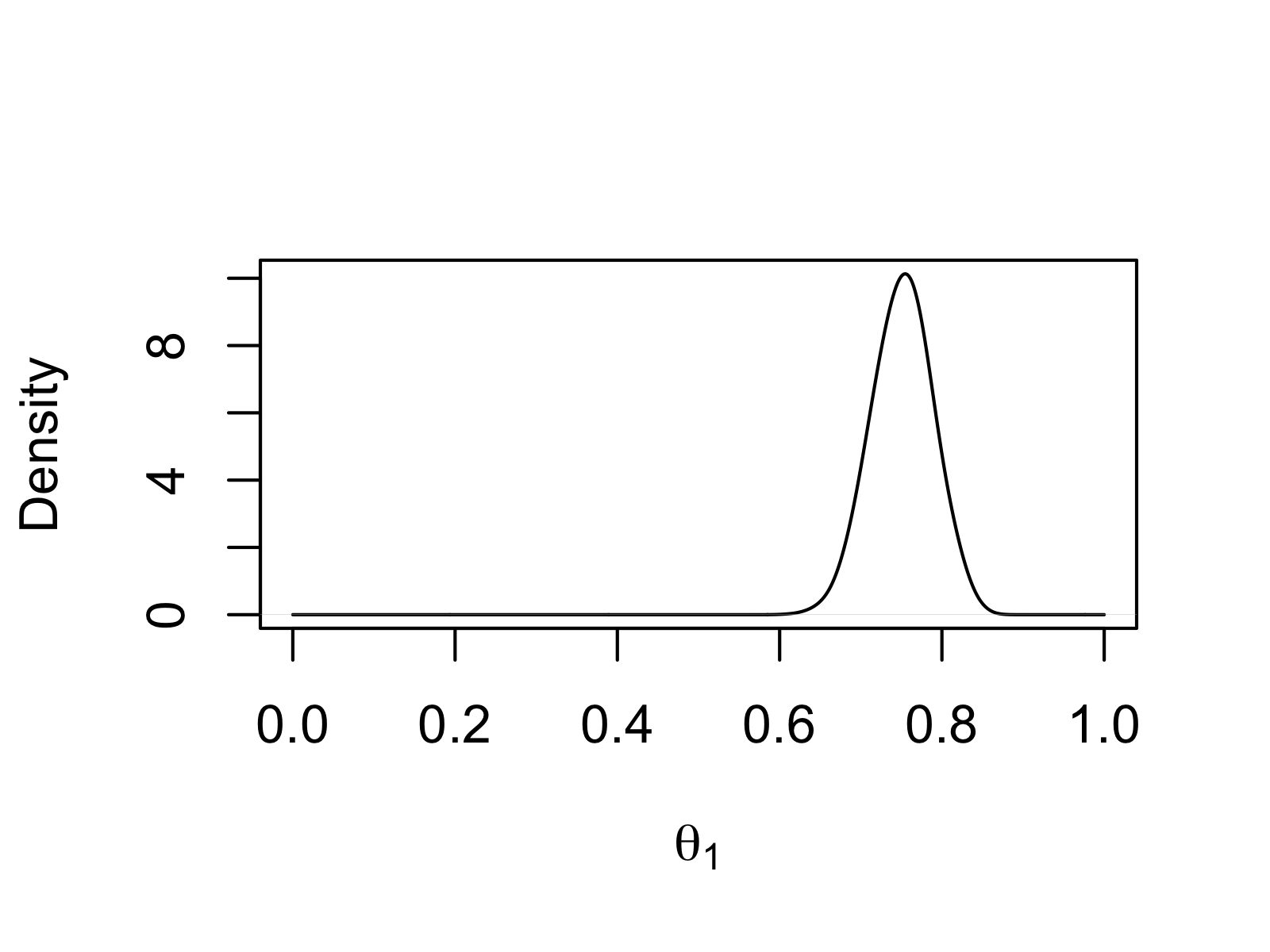}
	\caption{Density estimates of the posterior marginals for $\tau_1$, $\tau_2$, $\alpha$ and $\theta_1$,  from analysis of the measles data with the TSIR model.}
	\label{fig:histograms_m5}
\end{figure}

Figure \ref{fig:observed_fitted} shows the fit, and  it appears reasonable.
Figure \ref{fig:histograms_m5} displays  posterior distributions for a few interesting parameters. Notably, the posterior for $\alpha$ is close to uniform, so that the data tell us little about this parameter. The posterior for $\theta_1$ is quite peaked and the decay corresponding to this posterior is a swift decrease with increasing distance; $\tau_1$ and $\tau_2$ are relatively well-behaved. 


\subsection{Epidemic/Endemic Model}

The epidemic/endemic model we fit is,
\begin{align*}
Y_{it}|\mu_{it} & \sim \mbox{NegBin}(\mu_{it},\phi), \\
\mu_{it} & = \mbox{e}^{\lambda^{\AR}+ b_i^{\AR}} y_{i, t-1} 
+ \mbox{e}^{\lambda^{\NE} + b_i^{\NE}}\sum_{j=1}^n w_{ij} y_{j, t-1} +  N_{i} \mbox{e}^{\lambda^{\EN}_t+ b_i^{\EN}},
\end{align*}
where,
$$
w_{ij} = \frac{m_{ij}^{-\theta_2/(1-\theta_2)}}{\sum_{k \ne i} m_{ik}^{-\theta_2/(1-\theta_2)}},$$
with $m_{ij}$ the number of boundaries to cross when traveling between areas $i$ and $j$.
Independent normal random effects were used, i.e.,~$b_{i}^{\AR}  \sim N(0, \sigma^2_{\AR}) $, $b_{i}^{\NE}  \sim N(0, \sigma^2_{\NE}) $, $b_{i}^{\EN}  \sim N(0, \sigma^2_{\EN}) $.
Seasonality and a linear trend were included in the endemic component:
$$\lambda^{\EN}_t= \beta_0^{\EN}+ \beta_1^{\EN} t + \gamma \sin(\omega t) + \delta \cos(\omega t).$$
We assume relatively flat priors, as for the TSIR model, on $\lambda^{\AR}$, $\lambda^{\NE}$, $\beta_0^{\EN}$, $ \beta_1^{\EN}$, $\gamma$ and $\delta$. For the random effects precisions $\sigma^{-2}_{\AR}$,  $\sigma^{-2}_{\NE}$,  $\sigma^{-2}_{\EN}$ we use Gamma(0.5,0.1) priors which have prior medians for the standard deviations of 0.66, and (5\%, 95\%) points of (0.22, 7.1). We used a uniform prior on $0<\theta_2<1$ and a relatively flat prior on the overdispersion parameter $\phi$.

\begin{figure}[htbp]
	\centering
	\includegraphics[scale=0.09]{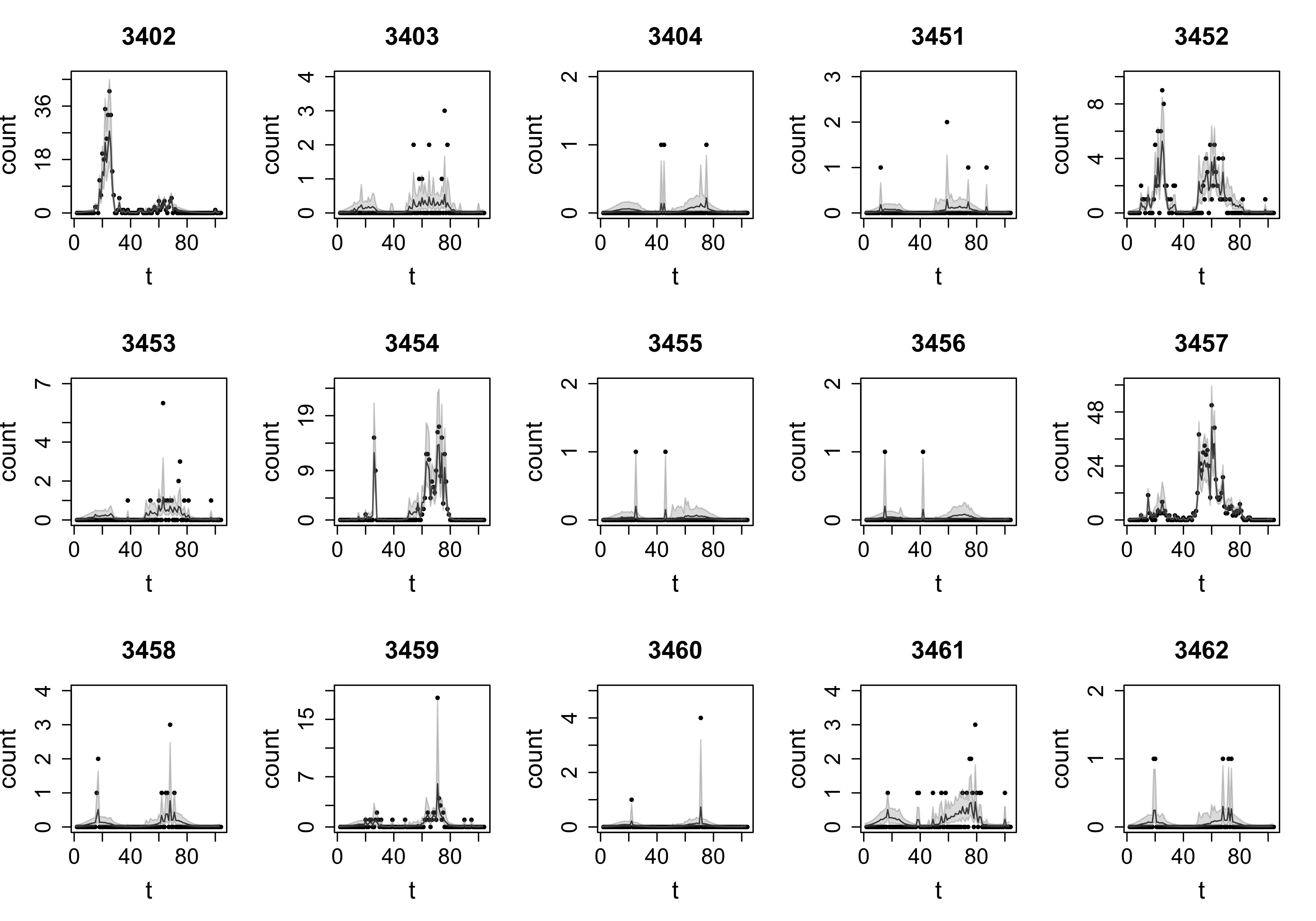}
	\caption{Observed (black dots) data in the 15 districts with non-zero counts, and posterior summaries (2.5\%, 50\%, 97.5\% quantiles) for $\mu_{it}$, under the epidemic/endemic model.}
	\label{fig:observed_fitted_hhh4}
\end{figure}

\begin{figure}[htbp]
	\centering
	\includegraphics[scale=0.09]{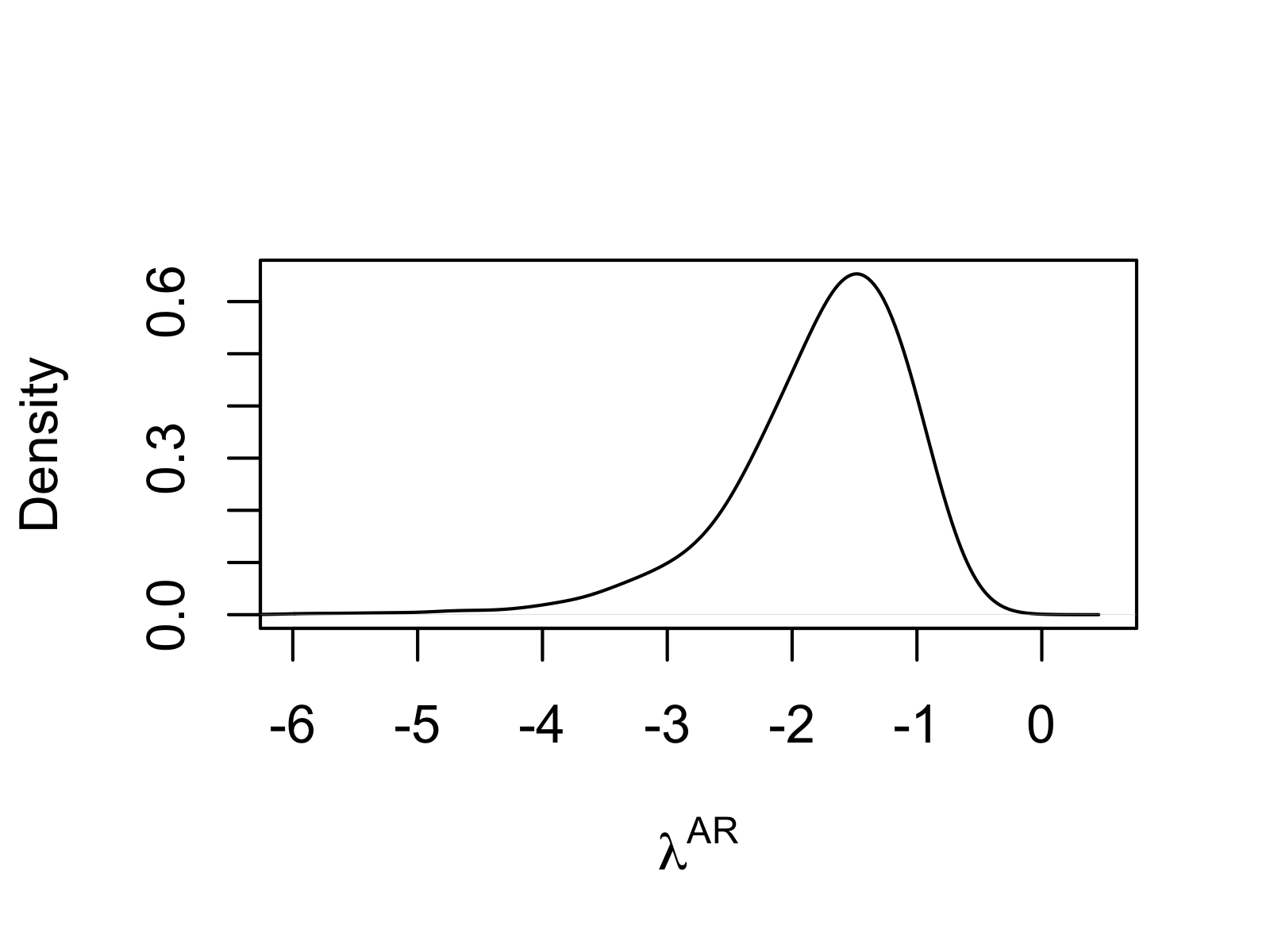}
	\includegraphics[scale=0.09]{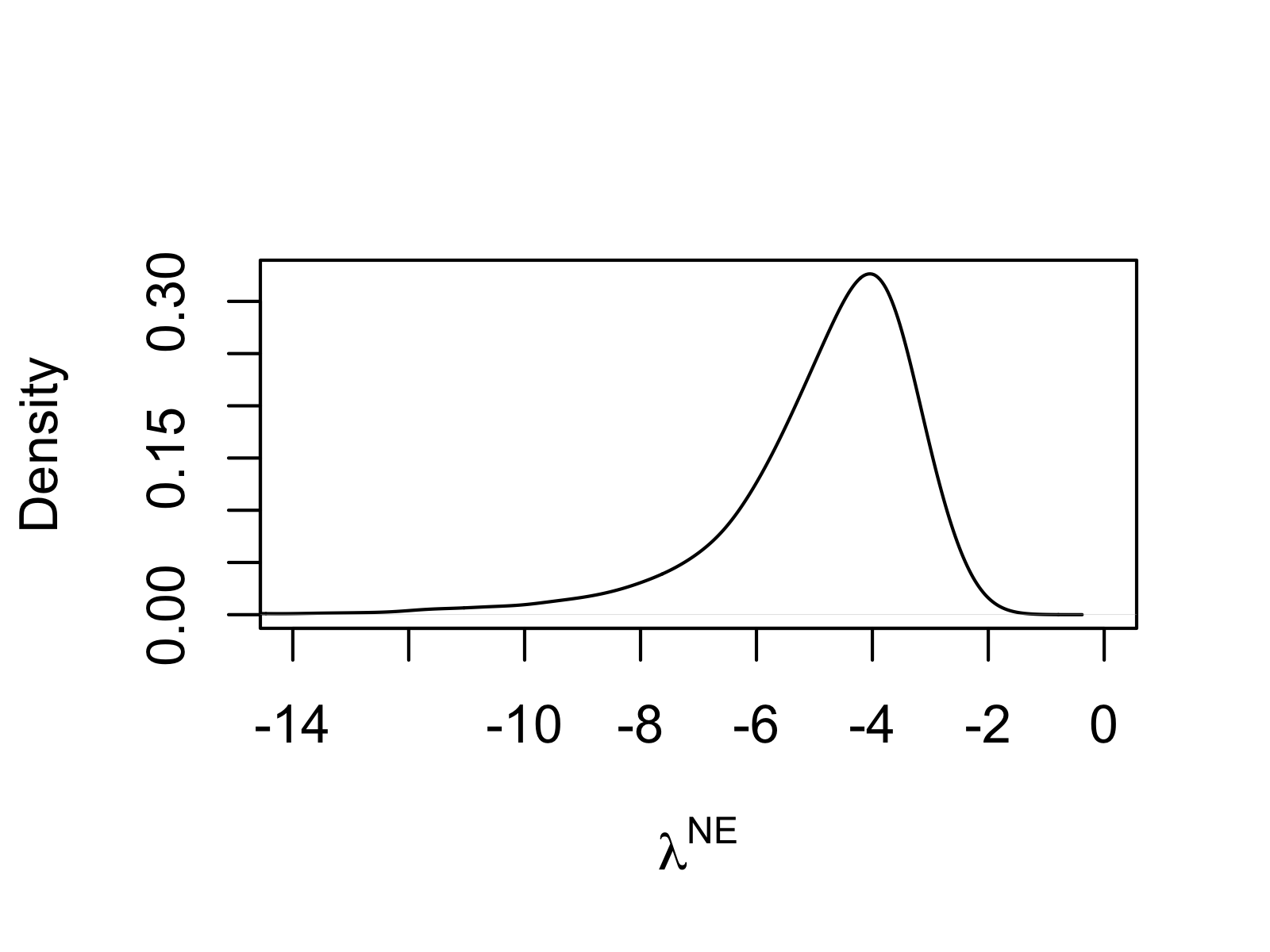}
	\includegraphics[scale=0.09]{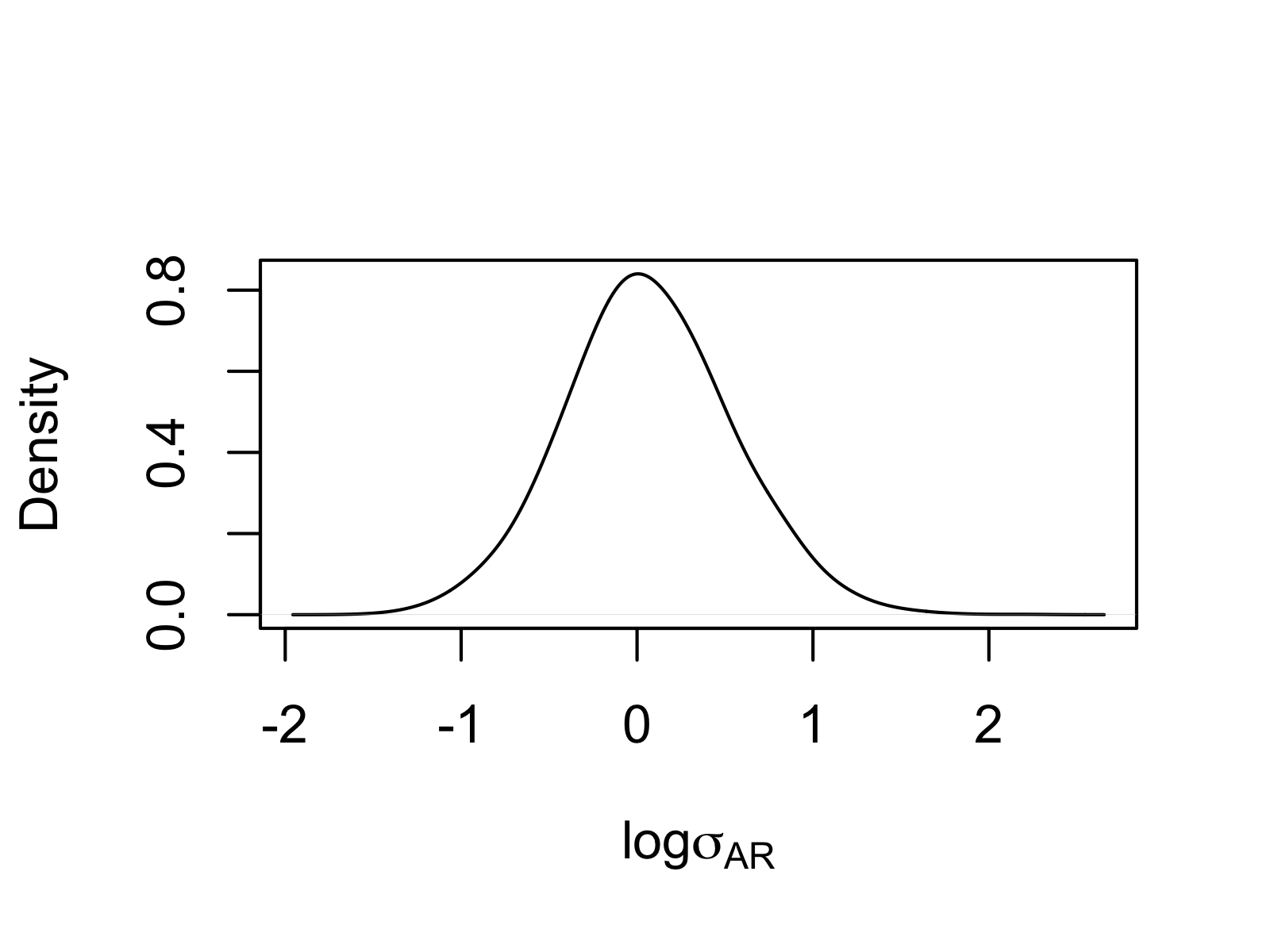}
	\includegraphics[scale=0.09]{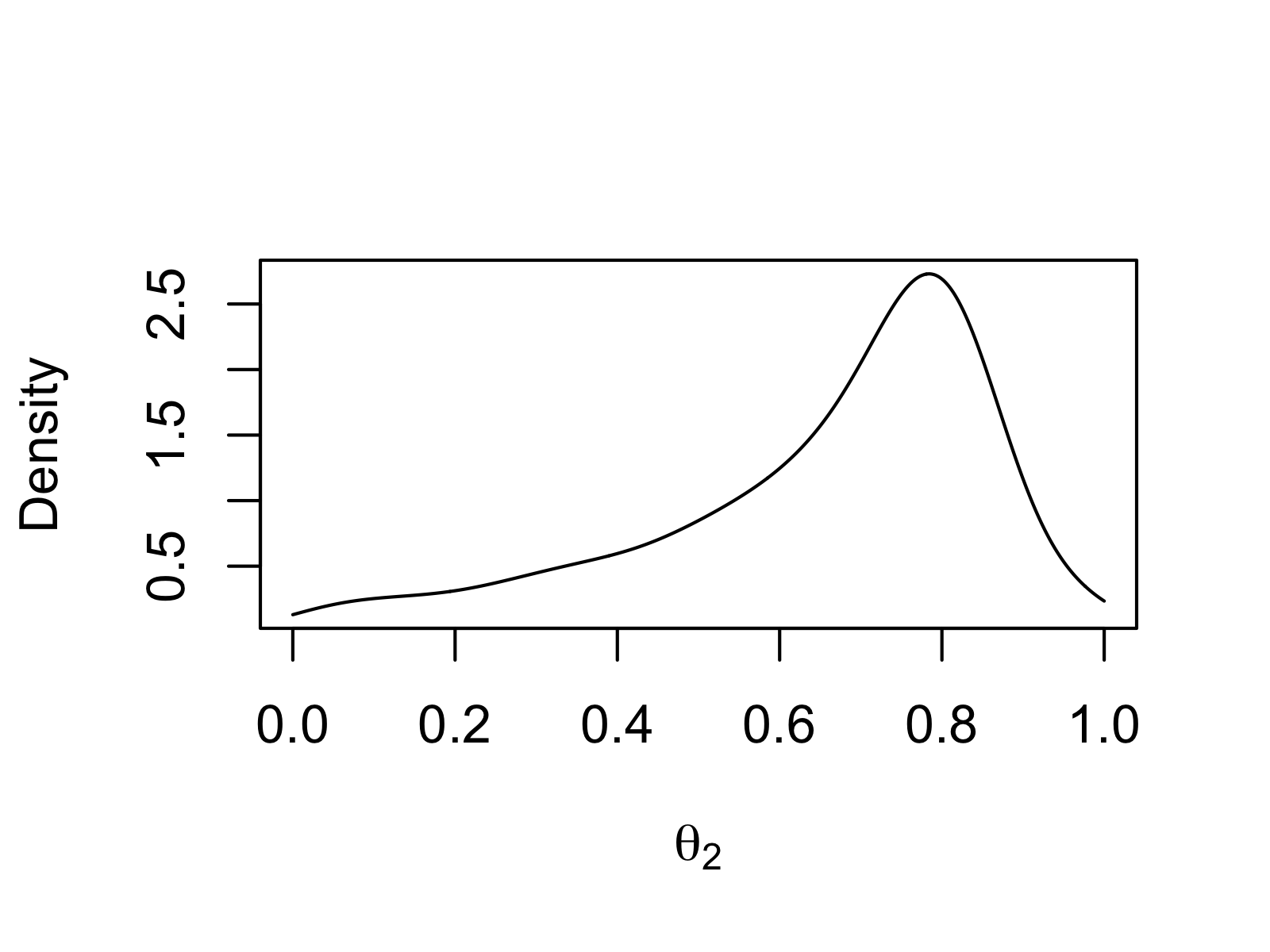}
	\caption{Density estimates of the posterior marginals for $\lambda^{\AR}$, $\lambda^{\NE}$,  $\log(\sigma_{\AR})$ and $\theta_2$, from analysis of the measles data with the epidemic-endemic model. 
	}
	\label{fig:hhh4_stan_m3/histogram_m3a}
\end{figure}

\begin{figure}[htbp]
	\centering
	\includegraphics[scale=0.15]{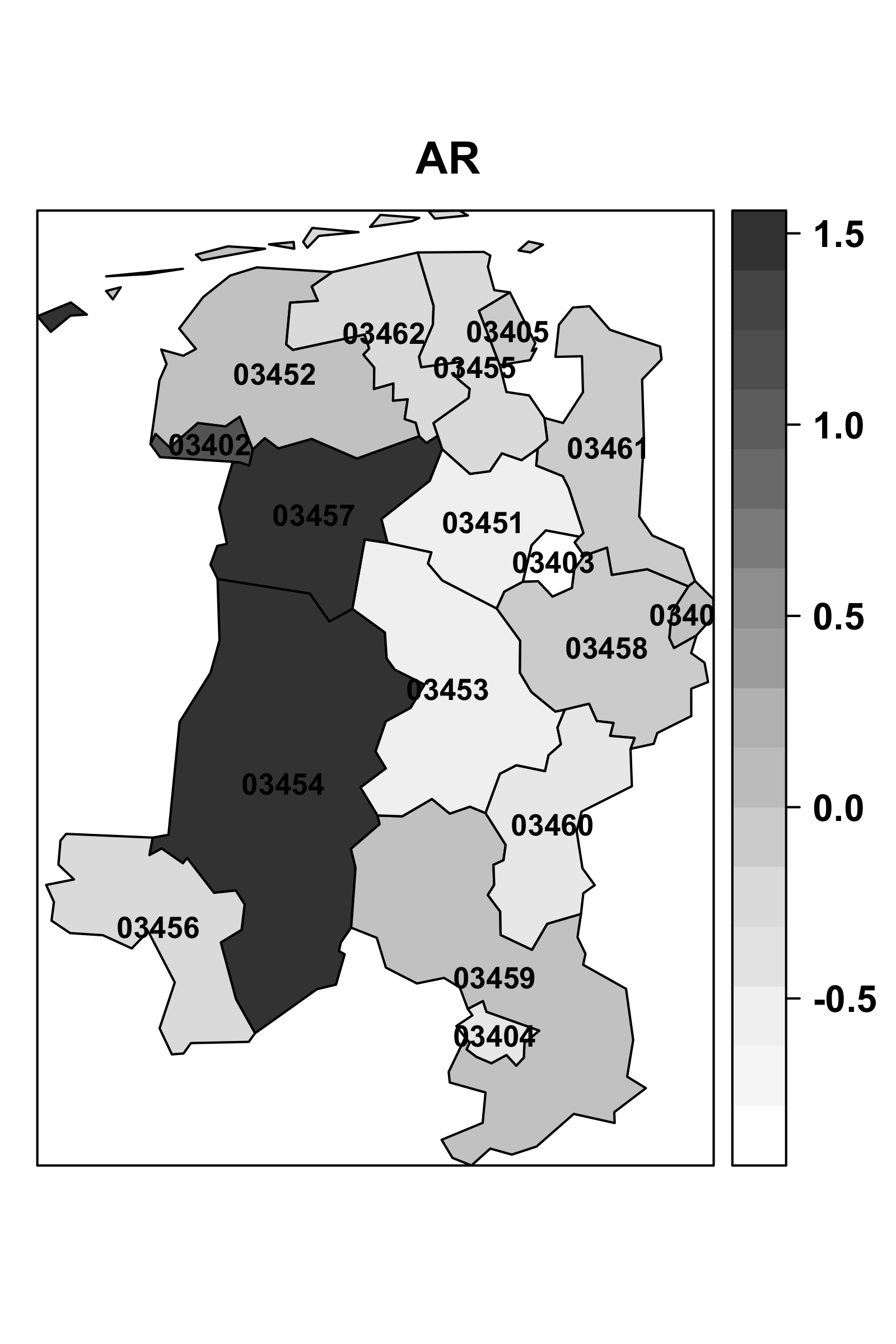}
	\caption{Posterior medians of autoregressive random effects $b_i^{\AR}$.}
	\label{fig:re_map}
\end{figure}

Figure \ref{fig:observed_fitted_hhh4} shows the time series of data in the areas with non-zero counts, along with posterior medians of $\mu_{it}$, with the gray shading representing 95\% intervals for $\mu_{it}$. Overall, the model appears to be picking up the time courses well.
Figure \ref{fig:hhh4_stan_m3/histogram_m3a} displays posterior distributions for a variety of parameters (for illustration), and we see that $\lambda^{\AR}$ and $\lambda^{\NE}$ are left-skewed. The neighborhood parameter $\theta_2$, is relatively well estimated and favors a nearest neighbor structure. The posterior medians of $\sigma_{\AR}$ and $\sigma_{\EN}$ were 1.5 and 1.6, respectively (though note that the random effects associated with these standard deviations are acting on different scales, so they are not comparable), while the posterior median of $\sigma_{\NE}$ is 3.3. Posterior medians of the random effects $b_i^{\AR}$ are mapped in Figure \ref{fig:re_map}, and we see large differences between the areas. In general, the sizes of the random effects show there is large between-area variability.

\section{Discussion}\label{sec:discussion}

In this review article we have concentrated on statistical aspects of the models, but perhaps the most difficult part is developing models and interpreting results in order to answer  biological questions. Much remains to be done on continuous time models, particularly with respect to computation. 

In the measles example, we illustrated that both TSIR and epidemic/endemic models can be fitted using {\tt Stan}. Residual analysis is under-developed for the discrete-time models that we have described, and model comparison in general also requires investigation. We examined conventional Pearson residuals in our analyses, but these are difficult to interpret with small counts. 
For underreporting, introducing the true counts as auxiliary variables is the obvious approach within a Bayesian model, but for large populations this is computationally very difficult (and also not currently possible in {\tt Stan} since it does not support inference for discrete parameters.

Contact patterns and disease severity are strongly age dependent, and incorporating age (and gender) in the epidemic/endemic model has been explicitly carried out \cite{bauer:wakefield:17} in the context of modeling hand, foot and mouth disease in China.   \cite{meyer:held:17} constructed contact matrices based on survey data and incorporated these  in the epidemic/endemic framework. Similar approaches would be straightforward to include in the TSIR approach.

\subsection*{Acknowledgements}

The authors would like to thank Bryan Grenfell and Saki Takahashi for helpful discussions on the TSIR model. Jon Wakefield was supported by grant R01 CA095994 from the National Institutes of Health, Tracy Qi Dong by grant U54 GM111274  from the National Institutes of Health and 
Vladimir Minin by grant R01 AI107034 from the National Institutes of Health.

\bibliographystyle{plain}
\bibliography{spatepi}

\begin{thebibliography}{10}

\bibitem{anderson:may:91}
R.M. Anderson and R.M. May.
\newblock {\em Infectious Diseases of Humans: Dynamics and Control}.
\newblock Oxford University Press, 1991.

\bibitem{bauer:wakefield:17}
C.~Bauer and J.~Wakefield.
\newblock Stratified space-time infectious disease modeling: with an
  application to hand, foot and mouth disease in {C}hina.
\newblock {\em Under revision}, 2017.

\bibitem{becker2017tsir}
A.D. Becker and B.T. Grenfell.
\newblock tsi{R}: An {R} package for time-series
  {S}usceptible-{I}nfected-{R}ecovered models of epidemics.
\newblock {\em PloS ONE}, 12:e0185528, 2017.

\bibitem{begon:etal:02}
M.~Begon, M.~Bennett, R.G. Bowers, N.P. French, S.M. Hazel, J.~Turner, et~al.
\newblock A clarification of transmission terms in host-microparasite models:
  numbers, densities and areas.
\newblock {\em Epidemiology and infection}, 129:147--153, 2002.

\bibitem{bjornstad:etal:02}
O.N. Bj{\o}rnstad, B.F. Finkenst{\"a}dt, and B.T. Grenfell.
\newblock Dynamics of measles epidemics: estimating scaling of transmission
  rates using a time series {SIR} model.
\newblock {\em Ecological Monographs}, 72:169--184, 2002.

\bibitem{bjornstad:grenfell:08}
O.N. Bj{\o}rnstad and B.T. Grenfell.
\newblock Hazards, spatial transmission and timing of outbreaks in epidemic
  metapopulations.
\newblock {\em Environmental and Ecological Statistics}, 15:265--277, 2008.

\bibitem{cardinal:etal:99}
M.~Cardinal, R.~Roy, and J.~Lambert.
\newblock On the application of integer-valued time series models for the
  analysis of disease incidence.
\newblock {\em Statistics in Medicine}, 18:2025--2039, 1999.

\bibitem{ciupe:etal:06}
M.S. Ciupe, B.L. Bivort, D.M. Bortz, and P.W. Nelson.
\newblock Estimating kinetic parameters from {HIV} primary infection data
  through the eyes of three different mathematical models.
\newblock {\em Mathematical biosciences}, 200:1--27, 2006.

\bibitem{cox:miller:65}
D.R. Cox and H.D. Miller.
\newblock {\em The Theory of Stochastic Processes}.
\newblock Chapman and Hall, 1965.

\bibitem{daley:gani:99}
D.J. Daley and J.~Gani.
\newblock {\em Epidemic Modelling: An Introduction}.
\newblock Cambridge University Press, 1999.

\bibitem{earn:etal:00}
D.J.D. Earn, P.~Rohani, B.M. Bolker, and B.T. Grenfell.
\newblock A simple model for complex dynamical transitions in epidemics.
\newblock {\em Science}, 287:667--670, 2000.

\bibitem{feller:50}
W.~Feller.
\newblock {\em An Introduction to Probability Theory and its Applications,
  Volume I}.
\newblock John Wiley \& Sons, 1950.

\bibitem{ferland:etal:06}
R.~Ferland, A.~Latour, and D.~Oraichi.
\newblock Integer-valued {GARCH} process.
\newblock {\em Journal of Time Series Analysis}, 27:923--942, 2006.

\bibitem{fernandez:etal:16}
A.~Fern{\'a}ndez-Fontelo, A.~Caba{\~n}a, P.~Puig, and D.~Mori{\~n}a.
\newblock Under-reported data analysis with {INAR}-hidden {M}arkov chains.
\newblock {\em Statistics in Medicine}, 35:4875--4890, 2016.

\bibitem{finkenstadt:grenfell:00}
B.F. Finkenst{\"a}dt and B.T. Grenfell.
\newblock Time series modelling of childhood diseases: a dynamical systems
  approach.
\newblock {\em Journal of the Royal Statistical Society: Series C},
  49:187--205, 2000.

\bibitem{fintzi:etal:17}
J.~Fintzi, X.~Cui, J.~Wakefield, and V.N. Minin.
\newblock Efficient data augmentation for fitting stochastic epidemic models to
  prevalence data.
\newblock {\em Journal of Computational and Graphical Statistics}, 2017.
\newblock To appear.

\bibitem{fisher:wakefield:17}
L.~Fisher and J.~Wakefield.
\newblock Ecological inference for infectious disease data, with application to
  vaccination strategies.
\newblock {\em In preparation}, 2017.

\bibitem{fokianos:fried:10}
K.~Fokianos and R.~Fried.
\newblock Interventions in {INGARCH} processes.
\newblock {\em Journal of Time Series Analysis}, 31:210--225, 2010.

\bibitem{geilhufe2014power}
M.~Geilhufe, L.~Held, S.O. Skr{\o}vseth, G.S. Simonsen, and F.~Godtliebsen.
\newblock Power law approximations of movement network data for modeling
  infectious disease spread.
\newblock {\em Biometrical Journal}, 56:363--382, 2014.

\bibitem{glass:etal:03}
K.~Glass, Y.~Xia, and B.T. Grenfell.
\newblock Interpreting time-series analyses for continuous-time biological
  models measles as a case study.
\newblock {\em Journal of Theoretical Biology}, 223:19--25, 2003.

\bibitem{greene:etal:16}
S.K. Greene, E.R. Peterson, D.~Kapell, A.D. Fine, and M.~Kulldorff.
\newblock Daily reportable disease spatiotemporal cluster detection, {N}ew
  {Y}ork {C}ity, {N}ew {Y}ork, {USA}, 2014--2015.
\newblock {\em Emerging Infectious Diseases}, 22:1808, 2016.

\bibitem{held:etal:05}
L.~Held, M.~H{\"o}hle, and M.~Hofmann.
\newblock A statistical framework for the analysis of multivariate infectious
  disease surveillance counts.
\newblock {\em Statistical Modelling}, 5:187--199, 2005.

\bibitem{held:paul:12}
L.~Held and M.~Paul.
\newblock Modeling seasonality in space-time infectious disease surveillance
  data.
\newblock {\em Biometrical Journal}, 54:824--843, 2012.

\bibitem{hoffman:gelman:14}
M.D. Hoffman and A.~Gelman.
\newblock The {N}o-{U}-turn sampler: adaptively setting path lengths in
  {H}amiltonian {M}onte {C}arlo.
\newblock {\em Journal of Machine Learning Research}, 15:1593--1623, 2014.

\bibitem{hohle:16}
M.~H{\"o}hle.
\newblock Infectious disease modeling.
\newblock In A.B. Lawson, S.~Banerjee, R.P. Haining, and M.D. Ugarte, editors,
  {\em Handbook of Spatial Epidemiology}, pages 477--500. Chapman and Hall/CRC
  Press, 2016.

\bibitem{hooker:etal:11}
G.~Hooker, S.P. Ellner, L.~De~Vargas~Roditi, and D.J.D. Earn.
\newblock Parameterizing state--space models for infectious disease dynamics by
  generalized profiling: measles in {O}ntario.
\newblock {\em Journal of The Royal Society Interface}, 8:961--974, 2011.

\bibitem{kendall:49}
D.G. Kendall.
\newblock Stochastic processes and population growth.
\newblock {\em Journal of the Royal Statistical Society, Series B},
  11:230--282, 1949.

\bibitem{kermack:mckendrick:27}
W.O. Kermack and A.G. McKendrick.
\newblock A contribution to the mathematical theory of epidemics.
\newblock {\em Proceedings of the Royal Society of London Series A, Containing
  Papers of a Mathematical and Physical Character}, 115:700--721, 1927.

\bibitem{koelle:pascual:04}
K.~Koelle and M.~Pascual.
\newblock Disentangling extrinsic from intrinsic factors in disease dynamics: a
  nonlinear time series approach with an application to cholera.
\newblock {\em The American Naturalist}, 163:901--913, 2004.

\bibitem{lekone:finkenstadt:06}
P.E. Lekone and B.F. Finkenst{\"a}dt.
\newblock Statistical inference in a stochastic epidemic {SEIR} model with
  control intervention: {E}bola as a case study.
\newblock {\em Biometrics}, 62:1170--1177, 2006.

\bibitem{levin:etal:15}
A.~Levin-Rector, E.L. Wilson, A.D. Fine, and S.K. Greene.
\newblock Refining historical limits method to improve disease cluster
  detection, {N}ew {Y}ork {C}ity, {N}ew {Y}ork, {USA}.
\newblock {\em Emerging Infectious Diseases}, 21:265, 2015.

\bibitem{liu:etal:87}
W.-M. Liu, H.W. Hethcote, and S.A. Levin.
\newblock Dynamical behavior of epidemiological models with nonlinear incidence
  rates.
\newblock {\em Journal of Mathematical Biology}, 25:359--380, 1987.

\bibitem{lloyd:etal:05}
J.O. Lloyd-Smith, S.J. Schreiber, P.E. Kopp, and W.M. Getz.
\newblock Superspreading and the effect of individual variation on disease
  emergence.
\newblock {\em Nature}, 438:355, 2005.

\bibitem{london:yorke:73}
W.P. London and J.A. Yorke.
\newblock Recurrent outbreaks of measles, chickenpox and mumps: I. seasonal
  variation in contact rates.
\newblock {\em American journal of epidemiology}, 98:453--468, 1973.

\bibitem{metcalf:etal:11rubellamexico}
C.J.E. Metcalf, O.N. Bj{\o}rnstad, M.J. Ferrari, P.~Klepac, N.~Bharti,
  H.~Lopez-Gatell, and B.T. Grenfell.
\newblock The epidemiology of rubella in {M}exico: seasonality, stochasticity
  and regional variation.
\newblock {\em Epidemiology and Infection}, 139:1029--1038, 2011.

\bibitem{meyer:held:14}
S.~Meyer and L.~Held.
\newblock Power-law models for infectious disease spread.
\newblock {\em Annals of Applied Statistics}, 8:1612--1639, 2014.

\bibitem{meyer:held:17}
S.~Meyer and L.~Held.
\newblock Incorporating social contact data in spatio-temporal models for
  infectious disease spread.
\newblock {\em Biostatistics}, 18:338--351, 2017.

\bibitem{meyer:etal:17}
S.~Meyer, L.~Held, and M.~H{\"o}hle.
\newblock Spatio-temporal analysis of epidemic phenomena using the {R} package
  {\tt surveillance}.
\newblock {\em Journal of Statistical Software}, 77, 2017.

\bibitem{morton:finkenstadt:05}
A.~Morton and B.F. Finkenst{\"a}dt.
\newblock Discrete time modelling of disease incidence time series by using
  {M}arkov chain {M}onte {C}arlo methods.
\newblock {\em Journal of the Royal Statistical Society, Series C},
  54:575--594, 2005.

\bibitem{neal:2011}
R.M. Neal.
\newblock {MCMC} using {H}amiltonian dynamics.
\newblock In S.~Brooks, A.~Gelman, G.L. Jones, and X.L. Meng, editors, {\em
  Handbook of Markov Chain Monte Carlo}, volume~2, pages 113--162. Chapman and
  Hall/CRC Press, 2011.

\bibitem{nishiura:etal:10}
H.~Nishiura, G.~Chowell, H.~Heesterbeek, and J.~Wallinga.
\newblock The ideal reporting interval for an epidemic to objectively interpret
  the epidemiological time course.
\newblock {\em Journal of The Royal Society Interface}, 7:297--307, 2010.

\bibitem{paul:held:11}
M.~Paul and L.~Held.
\newblock Predictive assessment of a non-linear random effects model for
  multivariate time series of infectious disease counts.
\newblock {\em Statistics in Medicine}, 30:1118--1136, 2011.

\bibitem{paul:etal:08}
M.~Paul, L.~Held, and A.~M. Toschke.
\newblock Multivariate modelling of infectious disease surveillance data.
\newblock {\em Statistics in Medicine}, 27:6250--6267, 2008.

\bibitem{prentice:etal:78}
R.L. Prentice, J.D. Kalbfleisch, A.V. Peterson~Jr, N.~Flournoy, V.T. Farewell,
  and N.E. Breslow.
\newblock The analysis of failure times in the presence of competing risks.
\newblock {\em Biometrics}, 34:541--554, 1978.

\bibitem{reich:etal:13}
N.G. Reich, S.~Shrestha, A.A. King, P.~Rohani, J.~Lessler, S.~Kalayanarooj,
  I.-K. Yoon, R.~V. Gibbons, D.~S. Burke, and D.A.T. Cummings.
\newblock Interactions between serotypes of dengue highlight epidemiological
  impact of cross-immunity.
\newblock {\em Journal of The Royal Society Interface}, 10:20130414, 2013.

\bibitem{takahashi:etal:16}
S.~Takahashi, Q.~Liao, T.P. Van~Boeckel, W.~Xing, J.~Sun, V.Y. Hsiao, C.~J.E.
  Metcalf, Z.~Chang, F.~Liu, J.~Zhang, J.T. Wu, B.J. Cowling, G.M. Leung, J.J.
  Farra, H.R. van Doorn, B.T. Grenfell, and H.~Yu.
\newblock Hand, foot, and mouth disease in {C}hina: modeling epidemic dynamics
  of enterovirus serotypes and implications for vaccination.
\newblock {\em PLoS Medicine}, 13:e1001958, 2016.

\bibitem{vanboeckel:etal:16}
T.P. Van~Boeckel, S.~Takahashi, Q.~Liao, W.~Xing, S.~Lai, V.~Hsiao, F.~Liu,
  Y.~Zheng, Z.~Chang, C.~Yuan, et~al.
\newblock Hand, foot, and mouth disease in {C}hina: critical community size and
  spatial vaccination strategies.
\newblock {\em Scientific Reports}, 6, 2016.

\bibitem{wakefield:08}
J.~Wakefield.
\newblock Ecologic studies revisited.
\newblock {\em Annual Review of Public Health}, 29:75--90, 2008.

\bibitem{wang:etal:11}
Y.~Wang, Z.~Feng, Y.~Yang, S.~Self, Y.~Gao, I.M. Longini, J.~Wakefield,
  J.~Zhang, L.~Wang, X.~Chen, L.~Yao, J.D. Stanaway, Z.~Wang, and W.~Yang.
\newblock Hand, foot and mouth disease in {C}hina: patterns and spread and
  transmissibility.
\newblock {\em Epidemiology}, 22:781--792, 2011.

\bibitem{xia:etal:04}
Y.~Xia, O.~N. Bj{\o}rnstad, and B.T. Grenfell.
\newblock Measles metapopulation dynamics: a gravity model for epidemiological
  coupling and dynamics.
\newblock {\em The American Naturalist}, 164:267--281, 2004.

\bibitem{zhu:11}
Fukang Zhu.
\newblock A negative binomial integer-valued {GARCH} model.
\newblock {\em Journal of Time Series Analysis}, 32:54--67, 2011.

\end{thebibliography}
%
\printindex

\end{document}